\documentclass[10pt,aps,prb,twocolumn,nofootinbib,floatfix]{revtex4-1}

\usepackage{amssymb,amsmath}
\usepackage{subfig}
\usepackage{graphicx}
\usepackage{color}
\usepackage{listings}
\usepackage{soul}
\usepackage[justification=Justified, singlelinecheck=off]{caption}
\usepackage{tikz}

\graphicspath{{figures/}}

\usepackage{natbib}
\begin{document}

\newcommand{\1}{{\bf \scriptstyle 1}\!\!{1}}
\newcommand{\unit}{\overleftrightarrow{{\bf \scriptstyle 1}\!\!{1}}}
\newcommand{\I}{{\rm i}}
\newcommand{\p}{\partial}
\newcommand{\D}{^{\dagger}}
\newcommand{\hbe}{\hat{\bf e}}
\newcommand{\bfa}{{\bf a}}
\newcommand{\bx}{{\bf x}}
\newcommand{\hbx}{\hat{\bf x}}
\newcommand{\by}{{\bf y}}
\newcommand{\hby}{\hat{\bf y}}
\newcommand{\br}{{\bf r}}
\newcommand{\hbr}{\hat{\bf r}}
\newcommand{\bj}{{\bf j}}
\newcommand{\bk}{{\bf k}}
\newcommand{\bn}{{\bf n}}
\newcommand{\bv}{{\bf v}}
\newcommand{\bp}{{\bf p}}
\newcommand{\bq}{{\bf q}}
\newcommand{\tp}{\tilde{p}}
\newcommand{\tbp}{\tilde{\bf p}}
\newcommand{\bu}{{\bf u}}
\newcommand{\hbz}{\hat{\bf z}}
\newcommand{\bA}{{\bf A}}
\newcommand{\calA}{\mathcal{A}}
\newcommand{\calB}{\mathcal{B}}
\newcommand{\tC}{\tilde{C}}
\newcommand{\bD}{{\bf D}}
\newcommand{\bE}{{\bf E}}
\newcommand{\calF}{\mathcal{F}}
\newcommand{\bB}{{\bf B}}
\newcommand{\bG}{{\bf G}}
\newcommand{\calG}{\mathcal{G}}
\newcommand{\obG}{\overleftrightarrow{\bf G}}
\newcommand{\bJ}{{\bf J}}
\newcommand{\bK}{{\bf K}}
\newcommand{\bL}{{\bf L}}
\newcommand{\tL}{\tilde{L}}
\newcommand{\bP}{{\bf P}}
\newcommand{\calP}{\mathcal{P}}
\newcommand{\bQ}{{\bf Q}}
\newcommand{\bR}{{\bf R}}
\newcommand{\bS}{{\bf S}}
\newcommand{\bH}{{\bf H}}
\newcommand{\balpha}{\mbox{\boldmath $\alpha$}}
\newcommand{\talpha}{\tilde{\alpha}}
\newcommand{\bsigma}{\mbox{\boldmath $\sigma$}}
\newcommand{\hbeta}{\hat{\mbox{\boldmath $\eta$}}}
\newcommand{\bSigma}{\mbox{\boldmath $\Sigma$}}
\newcommand{\bomega}{\mbox{\boldmath $\omega$}}
\newcommand{\bpi}{\mbox{\boldmath $\pi$}}
\newcommand{\bphi}{\mbox{\boldmath $\phi$}}
\newcommand{\hbphi}{\hat{\mbox{\boldmath $\phi$}}}
\newcommand{\btheta}{\mbox{\boldmath $\theta$}}
\newcommand{\hbtheta}{\hat{\mbox{\boldmath $\theta$}}}
\newcommand{\hbxi}{\hat{\mbox{\boldmath $\xi$}}}
\newcommand{\hbzeta}{\hat{\mbox{\boldmath $\zeta$}}}
\newcommand{\brho}{\mbox{\boldmath $\rho$}}
\newcommand{\bnabla}{\mbox{\boldmath $\nabla$}}
\newcommand{\bmu}{\mbox{\boldmath $\mu$}}
\newcommand{\bepsilon}{\mbox{\boldmath $\epsilon$}}

\newcommand{\iLambda}{{\it \Lambda}}
\newcommand{\cL}{{\cal L}}
\newcommand{\cH}{{\cal H}}
\newcommand{\cU}{{\cal U}}
\newcommand{\cT}{{\cal T}}

\newcommand{\be}{\begin{equation}}
\newcommand{\ee}{\end{equation}}
\newcommand{\bea}{\begin{eqnarray}}
\newcommand{\eea}{\end{eqnarray}}
\newcommand{\beqa}{\begin{eqnarray*}}
\newcommand{\eeqa}{\end{eqnarray*}}
\newcommand{\nn}{\nonumber}
\newcommand{\DD}{\displaystyle}

\newcommand{\ba}{\begin{array}{c}}
\newcommand{\baa}{\begin{array}{cc}}
\newcommand{\baaa}{\begin{array}{ccc}}
\newcommand{\baaaa}{\begin{array}{cccc}}
\newcommand{\ea}{\end{array}}

\newcommand{\bma}{\left[\begin{array}{c}}
\newcommand{\bmaa}{\left[\begin{array}{cc}}
\newcommand{\bmaaa}{\left[\begin{array}{ccc}}
\newcommand{\bmaaaa}{\left[\begin{array}{cccc}}
\newcommand{\ema}{\end{array}\right]}

\definecolor{dkgreen}{rgb}{0,0.6,0}
\definecolor{gray}{rgb}{0.5,0.5,0.5}
\definecolor{mauve}{rgb}{0.58,0,0.82}

\lstset{frame=tb,
  	language=Matlab,
  	aboveskip=3mm,
  	belowskip=3mm,
  	showstringspaces=false,
  	columns=flexible,
  	basicstyle={\small\ttfamily},
  	numbers=none,
  	numberstyle=\tiny\color{gray},
 	keywordstyle=\color{blue},
	commentstyle=\color{dkgreen},
  	stringstyle=\color{mauve},
  	breaklines=true,
  	breakatwhitespace=true
  	tabsize=3
}

\title{On-chip high-order parametric downconversion in the excitonic Mott insulator Nb$_3$Cl$_8$ for programmable multiphoton entangled states}
\author{Dmitry Skachkov}
\affiliation{NanoScience Technology Center, University of Central Florida, Orlando, FL USA}
\author{Dirk R. Englund}
\affiliation{Department of Electrical Engineering and Computer Science, Massachusetts Institute of Technology, Cambridge, MA 02139, USA.}
\author{Michael N. Leuenberger$^*$}
\affiliation{NanoScience Technology Center, Department of Physics, College of Optics and Photonics, University of Central Florida, Orlando, FL USA. \\ 
$^*$\textup{e-mail: Michael.Leuenberger@ucf.edu}}
\date{\today}

\begin{abstract}
Spontaneous parametric downconversion (SPDC) and four-wave mixing in $\chi^{(2)}$ and $\chi^{(3)}$ media underpin most entangled-photon sources, but direct generation of higher-order entangled multiphoton states by $n$-th order parametric downconversion remains extremely challenging because conventional materials exhibit tiny high-order nonlinearities. Here we show that single-layer Nb$_3$Cl$_8$, an excitonic Mott insulator on a breathing Kagome lattice, supports exceptionally large nonlinear susceptibilities up to seventh order. Many-body GW--Bethe--Salpeter and time-dependent BSE / Kadanoff--Baym simulations yield resonant $\chi^{(2)}$–$\chi^{(7)}$ for monolayer Nb$_3$Cl$_8$, with $|\chi^{(4)}|$ and $|\chi^{(5)}|$ surpassing values in prototypical transition metal dichalcogenides by 5–9 orders of magnitude. We trace this enhancement to flat bands and strongly bound Frenkel excitons with ferroelectrically aligned out-of-plane dipoles. Building on experimentally demonstrated 1$\times N$ integrated beam splitters with arbitrary power ratios, we propose an on-chip architecture where each output arm hosts an Nb$_3$Cl$_8$ patch, optionally gated by graphene to tune the complex $n$-photon amplitudes. Using the ab-initio $\chi^{(3)}$ and $\chi^{(4)}$ values, we predict that three-photon GHZ$_3$ and four-photon cluster-state sources in this platform can achieve $n$-photon generation rates up to $\sim 10^8$ and $\sim 10^6$ times larger, respectively, than silica-fiber- and MoS$_2$-based implementations with comparable geometry. We derive the quantum Hamiltonian and explicit $n$-photon generation rates for this platform, and show how suitable interferometric networks enable electrically and spectrally tunable GHZ, $W$, and cluster states based on genuine high-order nonlinear processes in a 2D excitonic Mott insulator.
\end{abstract}

\maketitle

\section{Introduction}

Entangled photons generated by spontaneous parametric downconversion (SPDC)\cite{Centini2005,Couteau2018,Hutter2020,Zeng2007,Liu2021,Slattery2019} in $\chi^{(2)}$ crystals and by spontaneous four-wave mixing\cite{Afsharnia2024,Cui2012,Wang2024_4} in $\chi^{(3)}$ media underpin a wide range of quantum technologies, including quantum key distribution,\cite{Pirandola2020,Zhang2025Q,Xu2020,Brazaola2024,Zapatero2023} quantum-enhanced metrology,\cite{Montenegro2025,Deng2024,Huang2024} and photonic quantum information processing.\cite{Slussarenko2019,Flamini2019,Delfanazari2025} Conventional sources rely on pairwise processes where a pump photon is converted into two lower-energy photons, or where two pump photons are converted into a signal--idler pair, and exploit bulk $\chi^{(2)}$ crystals, periodically poled waveguides, optical fibers, or nonlinear metasurfaces. Higher-order processes, such as third-order spontaneous parametric downconversion (3SPDC) producing entangled photon triplets, have been analyzed theoretically\cite{Domininguez2020,Bacaoco_TOPDC,Cavanna2020} and explored experimentally\cite{Corona_TOSPDC,Bertrand2025} in optical fibers, metasurfaces, and superconducting parametric cavities, but photon-triplet generation rates remain extremely small because $|\chi^{(3)}|$ and higher-order nonlinearities are weak in standard materials.\cite{Corona_TOSPDC,optical_fiber_TOSPDC,Chang_3photon_SPDC,Bacaoco_TOPDC}

The direct generation of $n$-photon entangled states by a single $n$-th order parametric process would provide an elegant path to Greenberger–Horne–Zeilinger (GHZ) states,\cite{Bao2024,Chen2024,Cao2024} $W$ states,\cite{Dur2000_Wclass,Eibl2004_expW,Menotti2016_energyW} Dicke states,\cite{Dicke1954,Stockton2003_Dicke,Toth2007_Dicke} non-Gaussian resource states,\cite{Walschaers2021,Kawasaki2024} and cluster states\cite{Nielsen2006,OSullivan2025,Freund2025,Lyubarov2025,Shi2021} for measurement-based quantum computing, without relying on probabilistic fusion of multiple pair sources. Realizing such sources in a compact, integrated platform requires materials that combine (i) large high-order nonlinear coefficients $\chi^{(n)}$, (ii) broadband phase-matching and dispersion engineering options, and (iii) compatibility with on-chip resonators or waveguides and programmable linear optics.

In parallel with progress on nonlinear materials, there has been rapid development of integrated 1$\times N$ power splitters and beam splitters with arbitrary splitting ratios. Early demonstrations based on topology-optimized or algorithmically designed nanostructures on silicon-on-insulator showed that arbitrary-ratio 1$\times 2$ and 1$\times 3$ splitters can be realized in footprints of only a few micrometers,\cite{Xu2017_OL,Xu2016_arxiv} while subsequent work introduced 1$\times N$ splitters for multimode platforms in which a non-uniform array of $N$ waveguides implements arbitrary power distributions among the outputs.\cite{Franz2021_JOpt} More recently, Ma \textit{et al.} used inverse design to obtain ultra-compact 1$\times N$ splitters with arbitrary input and output ports,\cite{Ma2020_SciRep} and Liu \textit{et al.} reported multiport arbitrary-ratio power splitters (MAPS) with three outputs on a silicon-on-insulator platform.\cite{Liu2025_Nanomaterials} Most relevant to the present work, Haines \textit{et al.}\cite{Haines2024_Nanophotonics} have experimentally fabricated and characterized 1$\times N$ integrated power splitters on silicon nitride that provide fully arbitrary splitting ratios for single- and multimode operation. Their devices are based on arrays of non-equally spaced waveguides and include, in particular, 1$\times 5$ splitters with ratios such as 1:1:2:1:1 at telecom wavelengths, providing a directly compatible, experimentally validated platform for distributing a pump field into $N$ coherent arms with programmable amplitudes.

Layered van der Waals materials have emerged as promising candidates for nonlinear and quantum photonics, due to their strong excitonic resonances, broken inversion symmetry in monolayers, and ease of integration with photonic structures. However, their nanometer-scale thickness often limits the effective nonlinear interaction length, and the reported high-harmonic generation (HHG) efficiencies and high-order nonlinear susceptibilities remain modest compared to what would be needed for practical $n$-photon sources.

In this work we identify single-layer Nb$_3$Cl$_8$ as a 2D material whose nonlinear response is strong enough to enable realistic $n$-th order parametric downconversion. Nb$_3$Cl$_8$ is a layered transition metal halide that realizes a breathing Kagome lattice of Nb atoms and hosts a correlated excitonic Mott insulating phase with a well-isolated flat electronic band structure.\cite{Sun_flatbands,Grytsiuk_Mott,Nakamura_HAXPES,Gao_flat_Mott} Our recent GW--BSE and time-dependent BSE (TD-BSE) simulations showed that monolayer Nb$_3$Cl$_8$ exhibits enormous nonlinear susceptibilities up to seventh order, with $\chi^{(4)}$ and $\chi^{(5)}$ exceeding those of MoS$_2$, GaSe, and other 2D semiconductors by several orders of magnitude.\cite{Khan_multiferroic_Nb3Cl8,Skachkov_GW_BSE_NL} These large nonlinearities arise from strongly bound Frenkel excitons localized on Nb$_3$ trimers, which carry permanent out-of-plane dipole moments aligned ferroelectrically due to the crystal field of the breathing Kagome lattice.\cite{Khan_multiferroic_Nb3Cl8}

Here we combine these \emph{ab initio} nonlinear coefficients with the toolbox of arbitrary-ratio 1$\times N$ integrated splitters to propose Nb$_3$Cl$_8$-based sources of entangled photon multiplets. In our architecture, a 1$\times N$ splitter (such as those of Haines \textit{et al.}\cite{Haines2024_Nanophotonics}) distributes the pump into $N$ arms, each hosting a monolayer Nb$_3$Cl$_8$ patch (optionally gated by graphene) that mediates $n$-th order spontaneous parametric downconversion with a complex amplitude $\kappa_\alpha$ tunable via local electrostatic control. We (i) summarize the excitonic and symmetry properties that underpin the nonlinear response of Nb$_3$Cl$_8$, (ii) present GW--BSE / TD-BSE results for $\chi^{(2)}$–$\chi^{(7)}$, and (iii) derive the effective $n$-photon creation Hamiltonian, explicit $n$-photon generation rates, and design rules for 1$\times N$ splitter-based devices that generate GHZ, $W$, and cluster states in realistic integrated geometries. Our analysis shows that the ultra-large $\chi^{(n)}$ values in Nb$_3$Cl$_8$ can compensate for its atomic thickness and, when combined with arbitrary-ratio 1$\times N$ splitters, enable programmable multiphoton entangled sources on a chip.

\section{Excitonic Mott insulator and ferroelectric Frenkel excitons in Nb$_3$Cl$_8$}

Nb\(_3\)Cl\(_8\) crystallizes in a layered structure in which Nb\(^{3+}\) ions with nominal \(d^2\) configuration form Nb\(_3\) trimers arranged in a breathing Kagome lattice within each layer, separated by Cl layers across van der Waals gaps (see Fig.~\ref{fig:Nb3Cl8_ML}).\cite{Sun_flatbands,Grytsiuk_Mott,Nakamura_HAXPES,Gao_flat_Mott} In the monolayer limit, the low-temperature structure belongs to the noncentrosymmetric point group \(C_{3v}\). The breathing distortion of the Kagome network leads to two inequivalent sets of Nb\(_3\) triangles and results in a spin-gapped ground state that can be described as a trimer-based Mott insulator.

Our GW--BSE calculations for monolayer Nb\(_3\)Cl\(_8\) reveal an isolated narrow band manifold near the Fermi energy and strongly bound Frenkel excitons localized on individual Nb\(_3\) trimers.\cite{Khan_multiferroic_Nb3Cl8} The lowest-energy exciton is a dark spin-triplet state with a binding energy exceeding the GW band gap and an excitation energy of approximately \(-0.5\)~eV relative to the electronic continuum. Its real-space wavefunction is confined to a single Nb\(_3\) cluster with an effective Bohr radius on the order of \(1.7\)~\AA, confirming its Frenkel character. By integrating the exciton charge density, we obtain a permanent electric dipole moment \(d \approx 0.73\)~D oriented nearly along the out-of-plane \(z\)-axis.\cite{Khan_multiferroic_Nb3Cl8}

The excitonic dipoles experience two competing interactions: (i) mutual dipole–dipole coupling between neighboring Nb\(_3\) trimers and (ii) coupling to the static crystal field generated by the ionic charges in Nb\(_3\)Cl\(_8\). Classical estimates using the ab-initio exciton density and Bader charges show that the crystal-field interaction stabilizing a ferroelectric out-of-plane alignment of the exciton dipoles is several orders of magnitude larger (on the scale of \(\sim 7\)~eV per Nb\(_3\) trimer) than the dipole–dipole interactions (meV per trimer). As a result, all Frenkel exciton dipoles are pinned ferroelectrically along the same out-of-plane direction, and the primitive triangular unit cell already captures the periodic dipole arrangement.

Because the monolayer structure has \(C_{3v}\) symmetry and lacks inversion, all even-order nonlinear susceptibilities \(\chi^{(2)}, \chi^{(4)}, \ldots\) are symmetry-allowed. In contrast, AB-stacked even-layer and bulk Nb\(_3\)Cl\(_8\) structures belong to the centrosymmetric group \(D_{3d}\), which forbids even-order nonlinear optical processes. The combination of: (i) strong inversion symmetry breaking in the monolayer, (ii) flat electronic bands, and (iii) localized, ferroelectrically aligned Frenkel excitons with a permanent dipole moment is the key ingredient responsible for the ultra-large nonlinear response of monolayer Nb\(_3\)Cl\(_8\).

\section{Ab initio evaluation of high-order nonlinear susceptibilities}
\label{sec:chi_abinitio}

To quantify the nonlinear optical response of monolayer Nb\(_3\)Cl\(_8\), we employ a combination of density functional theory (DFT), many-body GW, and Bethe–Salpeter equation (BSE) calculations for the equilibrium electronic and excitonic structure, followed by time-dependent BSE / Kadanoff–Baym simulations for the driven dynamics and extraction of high-order susceptibilities.\cite{Skachkov_GW_BSE_NL}

To compute the nonlinear susceptibilities, we solve the time-dependent BSE in the Kadanoff–Baym form for the nonequilibrium single-particle density matrix driven by a strong classical electromagnetic field. Two kinds of pump protocols are considered: (i) a monochromatic sinusoidal field switched on at \(t=0\), and (ii) a \(\delta\)-like ultrashort pulse at \(t=0\) followed, after 10~fs, by the sinusoidal pump. The latter protocol first prepares the system in the dark Frenkel exciton state and then probes its nonlinear response. The macroscopic polarization is extracted from the time-dependent density matrix, and the Fourier spectrum of the polarization is analyzed to obtain the harmonic components and, via appropriate normalization, the effective susceptibilities \(\chi^{(n)}(\omega)\) up to \(n=7\).\cite{Khan_multiferroic_Nb3Cl8,Skachkov_GW_BSE_NL}

Table~\ref{tab:chi} summarizes the resulting nonlinear coefficients at representative resonant photon energies and compares them to reported values for prototypical 2D materials and bulk nonlinear crystals. For monolayer Nb\(_3\)Cl\(_8\), we find
very large nonlinear optical coefficients,
with all values referring to the dominant in-plane tensor components.\cite{Khan_multiferroic_Nb3Cl8,Skachkov_GW_BSE_NL}

\begin{table}[t]
    \centering
    \caption{Nonlinear susceptibilities \(\chi^{(2)}\)–\(\chi^{(7)}\) for monolayer Nb\(_3\)Cl\(_8\) (this work) compared with representative experimental / theoretical values for other materials. Energies indicate representative photon energies where the corresponding response is evaluated.}
    \label{tab:chi}
    \begin{tabular}{lcccc}
    \hline\hline
    Material & Order & \(|\chi^{(n)}|\) & Units & \(E\) (eV) \\
    \hline
    Nb\(_3\)Cl\(_8\) & \(n=2\) & \(6.0\times 10^{-11}\) & m/V & 0.89 \\
                    & \(n=3\) & \(3.4\times 10^{-18}\) & (m/V)\(^2\) & 0.30 \\
                    & \(n=4\) & \(2.2\times 10^{-24}\) & (m/V)\(^3\) & 0.30 \\
                    & \(n=5\) & \(1.6\times 10^{-30}\) & (m/V)\(^4\) & 0.70 \\
                    & \(n=6\) & \(6.8\times 10^{-33}\) & (m/V)\(^5\) & 0.10 \\
                    & \(n=7\) & \(1.6\times 10^{-42}\) & (m/V)\(^6\) & 0.12 \\
    MoS\(_2\) monolayer & \(n=2\) & \(3\text{–}130\times 10^{-12}\) & m/V & \(\sim 0.8\text{–}1.2\) \\
                        & \(n=3\) & \(3.8\times 10^{-18}\) & (m/V)\(^2\) & 2.2 \\
                        & \(n=4\) & \(1.1\times 10^{-27}\) & (m/V)\(^3\) & 2.75 \\
                        & \(n=5\) & \(6.6\times 10^{-36}\) & (m/V)\(^4\) & 2.4 \\
    GaSe (few-layer)    & \(n=2\) & \(1.8\times 10^{-11}\) & m/V & \(\sim 1\) \\
                        & \(n=3\) & \(1.7\times 10^{-16}\) & (m/V)\(^2\) & \(\sim 1\) \\
    LiNbO\(_3\)         & \(n=2\) & \(2.6\times 10^{-11}\) & m/V & \(\sim 1\) \\
                        & \(n=3\) & \(2.0\times 10^{-25}\) & (m/V)\(^2\) & \(\sim 1\) \\
    \hline\hline
    \end{tabular}
\end{table}

We emphasize that for \(n\ge 4\) the Nb\(_3\)Cl\(_8\) susceptibilities surpass those of monolayer MoS\(_2\) and other 2D semiconductors by 5–9 orders of magnitude, and they exceed the corresponding bulk values of common nonlinear crystals such as LiNbO\(_3\), AlGaAs, and InP by several orders of magnitude at comparable photon energies.\cite{Boyd_NLO_review,MoS2_SHG_THz,LiNbO3_microresonator} This extreme enhancement is a direct manifestation of the flat-band excitonic Mott insulator physics and the ferroelectric exciton dipoles in Nb\(_3\)Cl\(_8\).
Details of the numerical calculations are shown in Sec.~\ref{sec:numerical}.

In the following sections, we use these ab initio susceptibilities as input parameters for a quantum-optical description of 
$n$-th order parametric downconversion in Nb$_3$Cl$_8$-based integrated photonic devices.

\begin{figure}[hbt]
\centering
\includegraphics[width=3.4in]{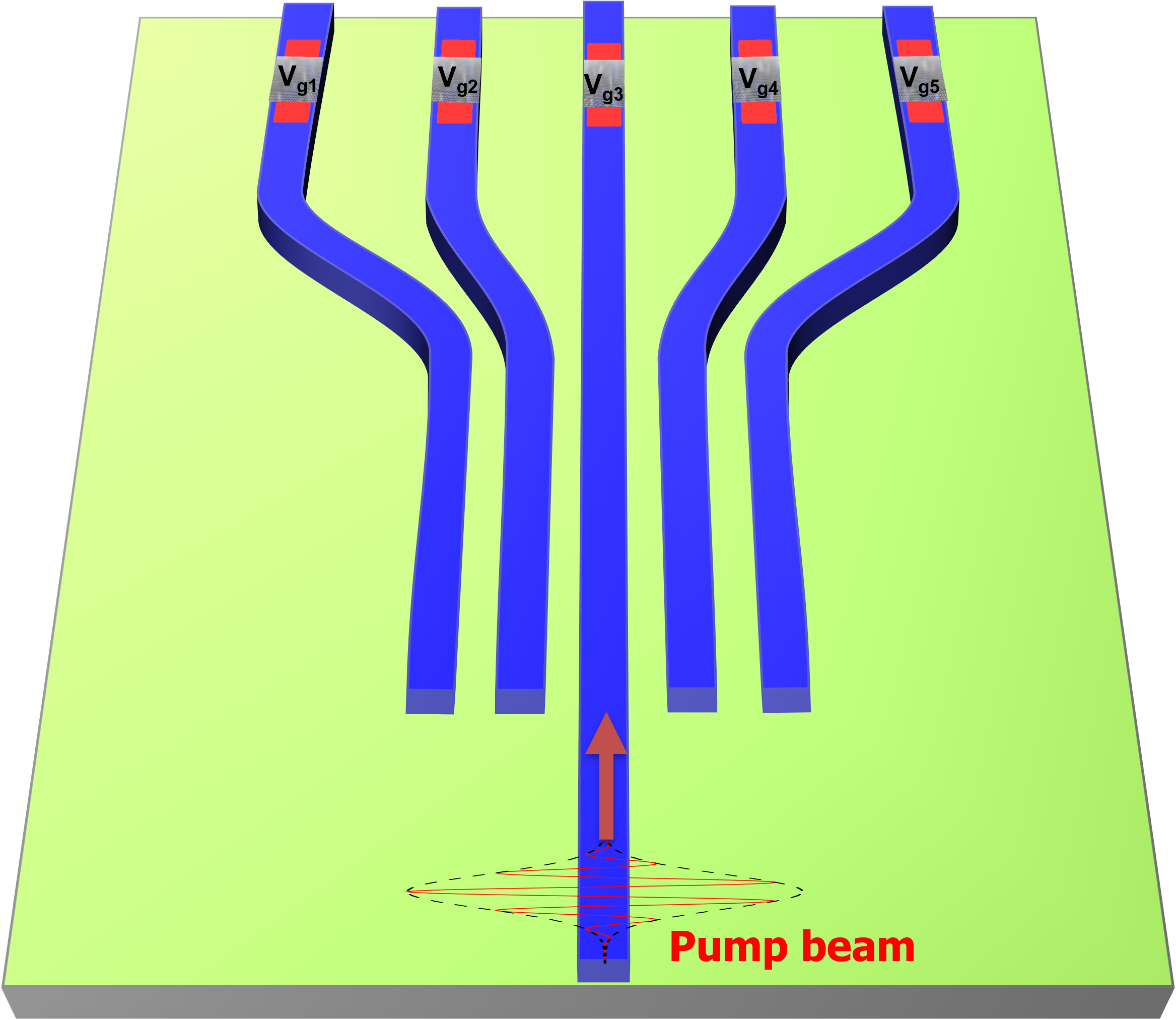}
\caption{Schematic of a 1$\times$5 Nb$_3$Cl$_8$--graphene high-order parametric downconversion device. A pump beam at frequency $\omega_p$ propagates in a single-mode input waveguide and is distributed by a 1$\times$5 beam splitter into five output waveguides. Each output arm contains a monolayer Nb$_3$Cl$_8$ patch acting as the high-order nonlinear medium, capped by a graphene gate biased at voltages $V_{g1},\dots,V_{g5}$. By tuning the gate voltages, the complex high-order susceptibility $\chi^{(n)}_{\mathrm{eff}}$ and local phase matching in each arm are modified, providing independent control over the complex $n$-photon downconversion amplitudes $\kappa_1,\dots,\kappa_5$ used to engineer multiphoton entangled states.}
\label{fig:1x5_beamsplitter}
\end{figure} 

\section{Quantum theory of $n$-th order parametric downconversion in a 2D sheet}
\label{sec:quantum_theory_PDC}

We now connect the large high-order susceptibilities of Nb$_3$Cl$_8$ to the quantum-optical process of $n$-th order parametric downconversion ($n$SPDC) in a 2D nonlinear sheet. We consider a classical pump field at frequency $\omega_p$ interacting with the monolayer and generating $n$ quantum modes at frequencies $\{\omega_j\}_{j=1,\dots,n}$ satisfying energy and momentum conservation,
\begin{equation}
\omega_p \approx \sum_{j=1}^{n} \omega_j, 
\qquad 
\mathbf{k}_p \approx \sum_{j=1}^{n} \mathbf{k}_j + \mathbf{G},
\end{equation}
where $\mathbf{G}$ is a reciprocal lattice vector accounting for quasi-phase-matching, periodic patterning, or a photonic crystal background.

The starting point is the nonlinear interaction Hamiltonian in the electric-dipole approximation,
\begin{equation}
H_{\text{int}}(t) = -\frac{\epsilon_0}{n!} \int dV\, 
\chi^{(n)}_{\alpha_1\ldots\alpha_n}(\mathbf{r};\{\omega\})\,
E_{\alpha_1}(\mathbf{r},t)\cdots E_{\alpha_n}(\mathbf{r},t),
\end{equation}
where repeated Cartesian indices are summed. The pump is treated as a classical field
\begin{equation}
\mathbf{E}_p(\mathbf{r},t) = \frac{1}{2}\left[\mathbf{E}_p(\mathbf{r})\,e^{-i\omega_p t} + \text{c.c.}\right],
\end{equation}
while the $n$ generated modes are quantized. For each generated frequency $\omega_j$ we write the quantized electric field as
\begin{equation}
\hat{\mathbf{E}}_j(\mathbf{r},t) 
= i \sum_{\mu_j} \sqrt{\frac{\hbar\omega_{j,\mu_j}}{2\epsilon_0 n_j^2 V_{j,\mu_j}}}
\left[\mathbf{f}_{j,\mu_j}(\mathbf{r})\,\hat{a}_{j,\mu_j} e^{-i\omega_{j,\mu_j} t} - \text{h.c.}\right],
\end{equation}
where $n_j$ is the effective refractive index for mode $j$, $V_{j,\mu_j}$ is an effective mode volume, $\mathbf{f}_{j,\mu_j}(\mathbf{r})$ is the normalized mode profile, and $\hat{a}_{j,\mu_j}$ is the annihilation operator for the corresponding cavity or waveguide mode.

Substituting these fields into $H_{\text{int}}$ and retaining only energy-conserving terms within the rotating-wave approximation yields, for a given set of modes, an effective $n$-photon creation Hamiltonian of the form
\begin{equation}
\hat{H}_{\text{PDC}} = \hbar \left[ \kappa_n\, \hat{a}_{1}^\dagger \hat{a}_{2}^\dagger \cdots \hat{a}_{n}^\dagger e^{-i\Delta t} + \text{h.c.} \right],
\label{eq:Hn}
\end{equation}
where $\Delta = \omega_p - \sum_{j} \omega_j$ is the detuning, and the effective coupling constant is
\begin{equation}
\kappa_n = \frac{\epsilon_0}{n!}\,\chi^{(n)}_{\text{eff}}\,E_p\,
\mathcal{F}_n
\prod_{j=1}^n \sqrt{\frac{\hbar\omega_j}{2\epsilon_0 n_j^2 V_j}}.
\label{eq:kappan}
\end{equation}
Here $\chi^{(n)}_{\text{eff}}$ is the relevant tensor component of $\chi^{(n)}$ for the chosen polarization configuration, $E_p$ is the pump field amplitude at the nonlinear sheet, and $\mathcal{F}_n$ is a dimensionless modal overlap integral,
\begin{equation}
\mathcal{F}_n 
= \frac{\displaystyle\int dV\, f_p(\mathbf{r})\, f_1^*(\mathbf{r})\cdots f_n^*(\mathbf{r})}{\displaystyle\sqrt{V_p V_1\cdots V_n}},
\end{equation}
which is naturally of order unity for well-confined and spatially overlapped modes. The dependence on the ab-initio nonlinear coefficients of Nb$_3$Cl$_8$ enters entirely through $\chi^{(n)}_{\text{eff}}$.

\subsection{Single-channel $n$-photon generation rate}

In the low-gain regime, where $|\kappa_n|T \ll 1$, the pump is undepleted and $\hat{H}_{\text{PDC}}$ can be treated as a weak perturbation. Starting from the vacuum of the $n$ output modes, $|i\rangle = |0\rangle$, and a given final state with one photon in each mode, $|f\rangle = |1_1,1_2,\ldots,1_n\rangle$, first-order time-dependent perturbation theory (equivalently, Fermi’s golden rule) gives the transition amplitude
\begin{equation}
c_f(T) \equiv \lambda_n(T)
= -\frac{i}{\hbar} \int_0^T dt\,
\langle f|\hat{H}_{\text{PDC}}(t)|i\rangle.
\end{equation}
Using Eq.~\eqref{eq:Hn} with
\begin{equation}
\hat{H}_{\text{PDC}}(t) 
= \hbar\left[\kappa_n
\hat{a}_1^\dagger\cdots \hat{a}_n^\dagger e^{-i\Delta t}
+ \text{h.c.}\right],
\end{equation}
and $\Delta = \omega_p - \sum_j \omega_j$, we obtain
\begin{align}
c_f(T) &= -\frac{i}{\hbar} \int_0^T dt\,
\hbar \kappa_n e^{-i\Delta t}
= -i \kappa_n \int_0^T dt\, e^{-i\Delta t}
\nonumber\\
&= -i \kappa_n \frac{1 - e^{-i\Delta T}}{\Delta}.
\end{align}
The corresponding transition probability is
\begin{align}
P_f(T) &= |c_f(T)|^2
= |\kappa_n|^2
\left|\frac{1 - e^{-i\Delta T}}{\Delta}\right|^2
\nonumber\\
&= |\kappa_n|^2 \frac{4\sin^2(\Delta T/2)}{\Delta^2}.
\end{align}
In the long–time limit $T \to \infty$, one uses the standard relation
\begin{equation}
\lim_{T\to\infty}
\frac{\sin^2(\Delta T/2)}{\pi T\,(\Delta/2)^2}
= 2\pi\,\delta(\Delta),
\end{equation}
so that the transition probability per unit time,
$R_n \equiv P_f(T)/T$, tends to
\begin{equation}
R_n = \frac{P_f(T)}{T}
\;\xrightarrow[T\to\infty]{}\;
2\pi |\kappa_n|^2 \delta(\Delta),
\end{equation}
which is just Fermi’s golden rule for this specific $n$-photon creation channel.

\subsection{Multi-mode output and effective $n$-photon density of states}

In a realistic cavity or waveguide, the output photons do not occupy a single discrete mode, but a set of modes labelled by quantum numbers $\mu_j$ (frequency, polarization, propagation direction, transverse index, path, etc.) for each channel $j=1,\dots,n$. The $n$-th order interaction Hamiltonian then generalizes to
\begin{equation}
\hat{H}_{\text{PDC}}(t)
= \hbar \sum_{\{\mu_j\}}
\left[
\kappa_n(\{\mu_j\})\,
\hat{a}_{1,\mu_1}^\dagger\cdots
\hat{a}_{n,\mu_n}^\dagger e^{-i\Delta(\{\mu_j\}) t}
+ \text{h.c.}
\right],
\end{equation}
where
\begin{equation}
\Delta(\{\mu_j\})
= \omega_p - \sum_{j=1}^n \omega_{j,\mu_j}
\end{equation}
is the detuning from energy conservation for the specific mode combination $\{\mu_j\}$, and $\kappa_n(\{\mu_j\})$ is the corresponding coupling constant.

For an initial vacuum state of all output modes $|i\rangle = |0\rangle$ and a final state with exactly one photon in each selected mode, $|f\rangle = |1_{1,\mu_1},\dots,1_{n,\mu_n}\rangle$, the first–order transition amplitude after an interaction time $T$ is
\begin{align}
c_f(T) &= -\frac{i}{\hbar} \int_0^T dt\,
\langle f|\hat{H}_{\text{PDC}}(t)|i\rangle
\nonumber\\
&= -i\,\kappa_n(\{\mu_j\})\,
\frac{1 - e^{-i\Delta(\{\mu_j\}) T}}{\Delta(\{\mu_j\})},
\end{align}
and the corresponding transition probability is
\begin{equation}
P_f(T) = |c_f(T)|^2
= |\kappa_n(\{\mu_j\})|^2
\frac{4\sin^2\!\left[\Delta(\{\mu_j\}) T/2\right]}
{\Delta^2(\{\mu_j\})}.
\end{equation}
The total $n$-photon generation probability is obtained by summing over all final states,
\begin{align}
P_{\text{tot}}(T)
&= \sum_{\{\mu_j\}} P_f(T)
\nonumber\\
&= \sum_{\{\mu_j\}}
|\kappa_n(\{\mu_j\})|^2
\frac{4\sin^2\!\left[\Delta(\{\mu_j\}) T/2\right]}
{\Delta^2(\{\mu_j\})}.
\end{align}
In the long–time limit $T\to\infty$, the same identity as above yields the total generation rate
\begin{equation}
R_n \equiv \frac{P_{\text{tot}}(T)}{T}
\xrightarrow[T\to\infty]{}
2\pi
\sum_{\{\mu_j\}}
|\kappa_n(\{\mu_j\})|^2
\delta\!\left[\omega_p - \sum_{j=1}^n \omega_{j,\mu_j}\right].
\label{eq:Rn_sum}
\end{equation}
For a waveguide or a leaky cavity, the set of modes is quasi-continuous and the discrete sums can be replaced by integrals over frequencies with the appropriate single-mode densities of states $\rho_j(\omega_j)$,
\begin{equation}
\sum_{\mu_j} \;\longrightarrow\;
\int d\omega_j\,\rho_j(\omega_j),
\qquad
\rho_n(\{\omega_j\}) \equiv \prod_{j=1}^n \rho_j(\omega_j),
\end{equation}
so that Eq.~\eqref{eq:Rn_sum} becomes
\begin{equation}
R_n = 2\pi \int
\left[\prod_{j=1}^n d\omega_j\,\rho_j(\omega_j)\right]
|\kappa_n(\{\omega_j\})|^2
\delta\!\left(\omega_p - \sum_{j=1}^n \omega_j\right).
\label{eq:Rn_int}
\end{equation}
Assuming that the joint coupling $|\kappa_n(\{\omega_j\})|^2$ varies slowly over the narrow bandwidth where the density of states and the $\delta$-function overlap, one can evaluate it at the central frequencies $\{\bar{\omega}_j\}$ that satisfy energy conservation, pull it out of the integral, and define an effective $n$-photon density of states $\rho_n^{\text{(eff)}}$,
\begin{equation}
\rho_n^{\text{(eff)}} \equiv
\int
\left[\prod_{j=1}^n d\omega_j\,\rho_j(\omega_j)\right]
|\mathcal{F}_n(\{\omega_j\})|^2\,
\delta\!\left(\omega_p - \sum_{j=1}^n \omega_j\right),
\label{eq:rho_n_eff}
\end{equation}
where we have included the modal overlap factor $|\mathcal{F}_n|^2$ in the definition for later convenience.

\subsection{Explicit $n$-photon rate in terms of $\chi^{(n)}$}

The microscopic expression for $\kappa_n(\{\mu_j\})$ follows from the nonlinear interaction Hamiltonian with the quantized signal modes and classical pump field. Using the usual cavity/waveguide normalization
\begin{equation}
\hat{\mathbf{E}}_{j}(\mathbf{r})
= i\sum_{\mu_j}
\sqrt{\frac{\hbar\omega_{j,\mu_j}}
{2\epsilon_0 n_j^2 V_{j,\mu_j}}}\,
\mathbf{f}_{j,\mu_j}(\mathbf{r})\,\hat{a}_{j,\mu_j}
+ \text{h.c.},
\end{equation}
and treating the pump field $\mathbf{E}_p(\mathbf{r})$ as classical, one finds
\begin{equation}
\kappa_n(\{\mu_j\})
= \frac{\epsilon_0}{n!}\,
\chi^{(n)}_{\text{eff}}\,E_p\,
\mathcal{F}_n(\{\mu_j\})
\prod_{j=1}^n
\sqrt{\frac{\hbar\omega_{j,\mu_j}}
{2\epsilon_0 n_j^2 V_{j,\mu_j}}},
\label{eq:kappan_multimode}
\end{equation}
where $\chi^{(n)}_{\text{eff}}$ is the relevant tensor component of the nonlinear susceptibility for the chosen polarization configuration, $E_p$ is the pump field amplitude at the nonlinear sheet, $V_{j,\mu_j}$ are effective mode volumes, and $\mathcal{F}_n(\{\mu_j\})$ is the modal overlap factor.

For clarity, we first compute
\begin{align}
|\kappa_n(\{\mu_j\})|^2
&=
\left(\frac{\epsilon_0}{n!}\right)^2
|\chi^{(n)}_{\text{eff}}|^2
|E_p|^2
|\mathcal{F}_n(\{\mu_j\})|^2
\prod_{j=1}^n
\frac{\hbar\omega_{j,\mu_j}}{2\epsilon_0 n_j^2 V_{j,\mu_j}}
\nonumber\\[4pt]
&=
\frac{\hbar^n\,\epsilon_0^{2-n}}{2^n (n!)^2}
|\chi^{(n)}_{\text{eff}}|^2
|E_p|^2
|\mathcal{F}_n(\{\mu_j\})|^2
\prod_{j=1}^n
\frac{\omega_{j,\mu_j}}{n_j^2 V_{j,\mu_j}}.
\label{eq:kappa_squared}
\end{align}
Inserting Eq.~\eqref{eq:kappa_squared} into Eq.~\eqref{eq:Rn_int}, and using the definition of $\rho_n^{(\text{eff})}$ in Eq.~\eqref{eq:rho_n_eff}, we obtain
\begin{equation}
R_n
= 2\pi\,\frac{\hbar^n\,\epsilon_0^{2-n}}{2^n (n!)^2}\,
|\chi^{(n)}_{\text{eff}}|^2\,|E_p|^2\,
\rho_n^{(\text{eff})}\,
\prod_{j=1}^n
\frac{\omega_j}{n_j^2 V_j},
\label{eq:Rn_full}
\end{equation}
where $\omega_j$, $n_j$, and $V_j$ are evaluated at the central frequencies of the participating modes. Equation~\eqref{eq:Rn_full} is the explicit form of the $n$-photon generation rate in a cavity or waveguide, including all constants and the effective $n$-photon density of states.

Neglecting order-unity variations in $\rho_n^{(\text{eff})}$ and $|\mathcal{F}_n|^2$, one can summarize this as the scaling relation
\begin{equation}
R_n
\sim
\frac{2\pi\,\hbar^n\,\epsilon_0^{2-n}}{2^n (n!)^2}\,
\rho_n^{(\text{eff})}\,
|\chi^{(n)}_{\text{eff}}|^2\,|E_p|^2
\prod_{j=1}^n
\frac{\omega_j}{n_j^2 V_j},
\label{eq:Rn_scaling}
\end{equation}
which makes explicit how $n$-photon downconversion is controlled by the magnitude of the high-order nonlinear susceptibility $\chi^{(n)}$, the pump field amplitude $E_p$, the mode volumes $V_j$, and the (geometry-dependent) effective $n$-photon density of states $\rho_n^{(\text{eff})}$.

For a 2D sheet of thickness $t$ much smaller than the wavelength, it is often convenient to work with an effective surface susceptibility $\chi^{(n)}_{\mathrm{surf}} = \chi^{(n)} t$ and to treat the monolayer as a nonlinear boundary condition in Maxwell's equations. In this picture, the effective interaction length is set by the cavity or waveguide length $L_{\text{cav}}$ and the sheet thickness enters through a factor $t/L_{\text{cav}}$, but the basic scaling with $\chi^{(n)}$ and the mode volumes remains unchanged.

Equation~\eqref{eq:Rn_scaling} immediately shows why high-order  processes have been so hard to realize in conventional materials: in standard dielectrics $|\chi^{(n)}|$ rapidly decreases with increasing $n$, and the resulting suppression of $R_n$ cannot be fully compensated by cavity enhancement alone. In contrast, the ultra-large $\chi^{(3)}-\chi^{(7)}$ of Nb$_3$Cl$_8$ obtained in Sec.~\ref{sec:chi_abinitio} indicate that $n$-photon downconversion with $n>2$ can become experimentally accessible in realistic integrated devices. In the following sections, we use Eq.~\eqref{eq:Rn_full} together with the ab-initio values of $\chi^{(n)}$ and concrete 1$\times N$ splitter geometries to estimate generation rates and to design architectures for multiphoton entangled state generation.

\section{Engineering multiphoton entangled states}
\label{sec:entangled_states}

The $n$-th order parametric interaction in monolayer Nb$_3$Cl$_8$ provides a resource for generating a wide variety of multiphoton entangled states. Which specific form of entanglement (GHZ, $W$, Dicke, cluster, \emph{etc.}) is obtained depends on: (i) the mode basis used to encode photonic qubits or qudits, (ii) the structure of the $n$-photon creation operators appearing in the effective Hamiltonian, (iii) the linear optical network following the nonlinear interaction, and (iv) any heralding or postselection applied in detection. In this section we outline how these ingredients can be combined, in conjunction with a 1$\times N$ splitter architecture, to engineer different classes of entangled states using the $n$-photon  process in Nb$_3$Cl$_8$.

\subsection{Mode basis, qubit encoding, and emission channels}
\label{subsec:mode_basis}

We assume that each of the $n$ photons generated in the  process is ultimately used to encode one photonic qubit (or qudit). A qubit can be encoded, for example, in
\begin{itemize}
  \item polarization: $|H\rangle_j \equiv \hat{a}_{H,j}^\dagger|0\rangle$, $|V\rangle_j \equiv \hat{a}_{V,j}^\dagger|0\rangle$,
  \item path (spatial mode): $|0\rangle_j \equiv \hat{a}_{A,j}^\dagger|0\rangle$, $|1\rangle_j \equiv \hat{a}_{B,j}^\dagger|0\rangle$,
  \item time bins: $|E\rangle_j \equiv \hat{a}_{E,j}^\dagger|0\rangle$, $|L\rangle_j \equiv \hat{a}_{L,j}^\dagger|0\rangle$,
  \item or frequency bins: $|\omega_{j}^{(0)}\rangle$, $|\omega_{j}^{(1)}\rangle$.
\end{itemize}
In all cases, the underlying field operators are linear combinations of the cavity or waveguide modes introduced in Sec.~\ref{sec:quantum_theory_PDC}. Formally, we can write a generic qubit basis as
\begin{equation}
|0\rangle_j = \hat{b}_{j,0}^\dagger |0\rangle,
\qquad
|1\rangle_j = \hat{b}_{j,1}^\dagger |0\rangle,
\end{equation}
where the $\hat{b}_{j,\ell}^\dagger$ are related to the physical cavity/waveguide modes $\hat{a}_{j,\mu}$ by some unitary transformation implemented by an on-chip linear optical network (beam splitters, phase shifters, polarizers, etc.). This freedom allows us to tailor the mapping between the \emph{natural} emission modes of the Nb$_3$Cl$_8$ device and the logical qubits used for quantum information processing.

In the low-gain regime, the $n$-photon state generated by the  interaction over a time $T$ can be written as
\begin{equation}
|\psi_{\text{out}}\rangle
\simeq
|0\rangle
- i \sum_{\alpha} \kappa_{\alpha} T\,\hat{A}_{\alpha}^\dagger |0\rangle,
\label{eq:psi_out_general}
\end{equation}
where $\alpha$ labels distinct emission \emph{channels},
\begin{equation}
\hat{A}_{\alpha}^\dagger
= \hat{a}_{1,\mu_1^{(\alpha)}}^\dagger
  \hat{a}_{2,\mu_2^{(\alpha)}}^\dagger
  \cdots
  \hat{a}_{n,\mu_n^{(\alpha)}}^\dagger,
\end{equation}
and $\kappa_{\alpha}$ is the corresponding (complex) coupling constant, determined by the overlap of the pump field, the Nb$_3$Cl$_8$ nonlinear polarization, and the particular set of output modes. By engineering which channels $\alpha$ are phase matched and how strongly they are driven, we can shape the superposition in Eq.~\eqref{eq:psi_out_general} and hence the resulting pattern of multiphoton entanglement.

In the 1$\times N$ architecture of Fig.~\ref{fig:1x5_beamsplitter}, a single pump mode at $\omega_p$ is distributed into $N$ spatially distinct arms by an integrated beam splitter with arbitrary power ratios.\cite{Haines2024_Nanophotonics,Franz2021_JOpt} Each arm contains a monolayer Nb$_3$Cl$_8$ patch (optionally gated by graphene), so that each branch realizes a separate $n$-photon emission channel with a tunable complex amplitude $\kappa_\alpha$. The arms thus provide a natural physical labeling of the channels $\alpha$.

\subsection{GHZ-type states from coherent superposition of arms}
\label{subsec:GHZ}

An $n$-qubit GHZ state has the form
\begin{equation}
|\text{GHZ}_n\rangle
=
\frac{1}{\sqrt{2}}\left(
|0,0,\ldots,0\rangle
+ e^{i\phi}\,|1,1,\ldots,1\rangle
\right),
\label{eq:GHZ_def}
\end{equation}
where each qubit is encoded as described above. To realize such a state directly from an $n$-th order  process, we require an interaction Hamiltonian that coherently couples the pump to two \emph{distinct} $n$-photon emission channels, both satisfying energy and momentum conservation, such that
\begin{equation}
\hat{H}_{\text{PDC}}^{(\text{GHZ})}
= \hbar\left[
\kappa_0 \hat{A}_0^\dagger
+ \kappa_1 e^{i\phi} \hat{A}_1^\dagger
+ \text{h.c.}
\right],
\label{eq:H_GHZ}
\end{equation}
with
\begin{equation}
\hat{A}_0^\dagger
= \hat{b}_{1,0}^\dagger
  \hat{b}_{2,0}^\dagger
  \cdots
  \hat{b}_{n,0}^\dagger,
\qquad
\hat{A}_1^\dagger
= \hat{b}_{1,1}^\dagger
  \hat{b}_{2,1}^\dagger
  \cdots
  \hat{b}_{n,1}^\dagger.
\label{eq:A0A1_def}
\end{equation}
Physically, $\hat{A}_0^\dagger$ and $\hat{A}_1^\dagger$ correspond to two different nonlinear conversion pathways that generate $n$ photons into two distinguishable sets of modes. In the 1$\times N$ geometry of Fig.~\ref{fig:1x5_beamsplitter}, these can be realized by two subsets of arms (e.g.\ arms 1 and 2) whose  outputs are subsequently mapped to the logical $|0\rangle_j$ and $|1\rangle_j$ modes of each qubit via a fixed interferometric network.

In the low-gain regime, Eq.~\eqref{eq:H_GHZ} yields (up to an irrelevant global phase)
\begin{equation}
|\psi_{\text{GHZ}}\rangle
\simeq
|0\rangle
- i T \left(
\kappa_0 \hat{A}_0^\dagger
+ \kappa_1 e^{i\phi} \hat{A}_1^\dagger
\right)|0\rangle.
\end{equation}
If the device is designed such that $|\kappa_0| = |\kappa_1|$ and the relative phase between the two pathways is controlled by graphene gates and integrated phase shifters, the (unnormalized) one-emission component is
\begin{equation}
|\psi_{\text{GHZ}}^{(1)}\rangle
\propto
\hat{A}_0^\dagger|0\rangle
+ e^{i\phi}\hat{A}_1^\dagger|0\rangle
=
|0,0,\ldots,0\rangle
+ e^{i\phi}|1,1,\ldots,1\rangle,
\end{equation}
which is precisely the GHZ state in Eq.~\eqref{eq:GHZ_def}. Thus, GHZ-type multiphoton states can be generated in a single emission event by
(i) enabling two phase-matched $n$-photon emission channels (two subsets of arms in the 1$\times N$ splitter) that differ only by the logical basis ($|0\rangle_j$ vs $|1\rangle_j$) for each qubit,
(ii) pumping them coherently with a well-defined relative phase,
(iii) and operating in the low-gain regime so that multi-pair events are negligible.

\subsection{$W$ and Dicke states from mode mixing and postselection}
\label{subsec:W_state}

The $n$-qubit $W$ state,
\begin{align}
|W_n\rangle
&= \frac{1}{\sqrt{n}}\left(
|1,0,0,\ldots,0\rangle
+ |0,1,0,\ldots,0\rangle
+ \cdots
\right.\nn\\
&+\left. |0,0,\ldots,1\rangle
\right),
\label{eq:W_def}
\end{align}
contains only a \emph{single} excitation delocalized among $n$ qubits, whereas the natural output of an $n$-th order  process is an $n$-photon Fock state $|1_1,1_2,\ldots,1_n\rangle$. To obtain $W$ or more general Dicke states from the $n$-photon resource, one combines two key steps:
(i) a linear optical network that mixes the $n$ physical modes $\hat{a}_j$ (arising from the different arms of the splitter) into $n$ logical modes $\hat{b}_{j,\ell}$,
(ii) heralding or postselection on specific detection patterns that project the $n$-photon state onto the desired single-excitation (or $m$-excitation) subspace.

Let $\hat{\boldsymbol{b}} = U \hat{\boldsymbol{a}}$ be an $n\times n$ unitary transformation implemented by an on-chip interferometer, where $\hat{\boldsymbol{a}} = (\hat{a}_1,\ldots,\hat{a}_n)^T$ collects the physical modes in the different arms and $\hat{\boldsymbol{b}} = (\hat{b}_1,\ldots,\hat{b}_n)^T$ the logical modes. The $n$-photon creation operator then transforms as
\begin{equation}
\hat{a}_1^\dagger \hat{a}_2^\dagger \cdots \hat{a}_n^\dagger
= \prod_{j=1}^n
\left(\sum_{k=1}^n U_{kj}^* \hat{b}_k^\dagger\right),
\end{equation}
which expands into a coherent superposition of terms with different photon-number distributions in the $\hat{b}_k$ modes. By choosing $U$ appropriately (e.g.\ a symmetric multiport beam splitter), one can arrange that the amplitude for finding exactly one photon spread over the $n$ logical modes,
\begin{equation}
\hat{b}_1^\dagger|0\rangle,
\quad
\hat{b}_2^\dagger|0\rangle,
\quad \dots,\quad
\hat{b}_n^\dagger|0\rangle,
\end{equation}
is symmetric, while additional heralding detectors in ancillary modes are used to discard events with unwanted photon-number patterns. Conditioned on the desired detection pattern, the remaining $n$ logical qubits can be projected onto the $W_n$ state in Eq.~\eqref{eq:W_def}.

More generally, symmetric Dicke states with $m$ excitations,
\begin{equation}
|D_n^{(m)}\rangle
= \binom{n}{m}^{-1/2}
\sum_{\text{perm}}
|1,\ldots,1,0,\ldots,0\rangle,
\end{equation}
can be realized by appropriately designing $U$ and the heralding scheme. In the context of Nb$_3$Cl$_8$, the large $n$-th order coupling $\kappa_n$ derived in Sec.~\ref{sec:quantum_theory_PDC} is crucial, because it enhances the probability of generating the initial $n$-photon Fock state from which $W$ and Dicke states are distilled.

\subsection{Cluster and graph states from multi-mode interactions}
\label{subsec:cluster_states}

One of the main applications of multiphoton entanglement is the generation of cluster and more general graph states for measurement-based quantum computation. A photonic cluster state on a graph $G$ with vertices corresponding to qubits and edges to controlled-phase (CZ) entangling operations can be written as
\begin{equation}
|C_G\rangle
=
\prod_{(j,k)\in E(G)}
\hat{U}_{\text{CZ}}^{(jk)}
\bigotimes_{j=1}^N
|+\rangle_j,
\quad
|+\rangle_j =
\frac{|0\rangle_j + |1\rangle_j}{\sqrt{2}},
\end{equation}
where $\hat{U}_{\text{CZ}}^{(jk)}$ is a CZ gate acting on qubits $j$ and $k$.

In an Nb$_3$Cl$_8$-based device, there are two complementary routes to cluster states:
\begin{enumerate}
  \item \emph{Direct $n$-photon graph states from engineered  Hamiltonians.}  
  By exploiting the rich mode structure of a 1$\times N$ splitter plus interferometric network, one can design a Hamiltonian that couples the pump to several distinct $n$-photon emission channels corresponding to different groupings of arms, e.g.
  \begin{equation}
  \hat{H}_{\text{PDC}}^{(\text{graph})}
  = i\hbar\sum_{\alpha\in \mathcal{E}}
  \kappa_{\alpha}
  \left(\hat{A}_{\alpha}^\dagger - \hat{A}_{\alpha}\right),
  \end{equation}
  where $\mathcal{E}$ is a set of ``hyperedges'' linking subsets of arms. After mapping these modes to logical qubits via a linear optical network, the resulting output state can approximate a graph state whose adjacency structure is encoded in the pattern of $\kappa_{\alpha}$.
  \item \emph{Fusion-based cluster state generation.}  
  Alternatively, one can use the $n$-photon  source as a generator of smaller resource states (such as GHZ$_n$ or $W_n$), which are then fused into larger cluster states using linear optics and photon detection. For example, two GHZ$_3$ states produced by successive  events can be fused into a 4-qubit linear cluster by interfering one photon from each triplet on a beam splitter and postselecting on specific coincidence detection events. The high brightness of the Nb$_3$Cl$_8$ source is essential here, as fusion protocols are typically probabilistic and benefit strongly from higher source rates.
\end{enumerate}

In both approaches, on-chip integration of Nb$_3$Cl$_8$ with reconfigurable interferometric meshes (e.g.\ Mach--Zehnder lattices) enables programmable control over the mode basis and effective interaction graph, allowing cluster states of different sizes and topologies to be generated on demand.

\subsection{Tunable engineering with graphene-integrated 1$\times N$ splitters}
\label{subsec:graphene_tunable}

So far we have treated the $n$-th order coupling constants $\kappa_{\alpha}$ in Eq.~\eqref{eq:psi_out_general} as fixed parameters set by the device geometry and material properties. In practice, it is highly desirable to make these couplings \emph{tunable}, in both magnitude and phase, so that different multiphoton entangled states (GHZ, $W$, Dicke, cluster, \emph{etc.}) can be selected or reconfigured \emph{in situ}. One promising route, illustrated in Fig.~\ref{fig:1x5_beamsplitter}, is to integrate monolayer Nb$_3$Cl$_8$ with graphene contacts on top of each arm of a 1$\times N$ splitter. By electrostatically gating the graphene, one can tune its complex conductivity $\sigma(\omega,E_F)$ and thereby modify: (i) the effective dielectric environment experienced by Nb$_3$Cl$_8$, (ii) the dispersion and confinement of the guided modes, and (iii) the excitonic resonances and high-order nonlinear coefficients of Nb$_3$Cl$_8$ via screening and band renormalization.

In a simple model, the complex $n$-th order coupling constant for a particular emission channel $\alpha$ (i.e.\ a specific arm or subset of arms) can be written as
\begin{equation}
\kappa_{\alpha}(V_g)
=
\kappa_{\alpha}^{(0)}
\,\mathcal{M}_{\alpha}(V_g)\,
\mathcal{P}_{\alpha}(V_g),
\label{eq:kappa_split}
\end{equation}
where $V_g$ is the applied gate voltage on the graphene contacts. The factor $\mathcal{M}_{\alpha}(V_g)$ collects all changes in the \emph{magnitude} of the nonlinear interaction,
\begin{equation}
\mathcal{M}_{\alpha}(V_g)
=
\frac{\chi^{(n)}_{\text{eff}}(V_g)}{\chi^{(n)}_{\text{eff}}(0)}\,
\frac{E_p(V_g)}{E_p(0)}\,
\frac{\mathcal{F}_n^{(\alpha)}(V_g)}{\mathcal{F}_n^{(\alpha)}(0)},
\end{equation}
where $\chi^{(n)}_{\text{eff}}(V_g)$ is the (complex) effective $n$-th order susceptibility of Nb$_3$Cl$_8$ in the gated environment, $E_p(V_g)$ is the intracavity pump field amplitude, and $\mathcal{F}_n^{(\alpha)}(V_g)$ is the modal overlap factor for channel $\alpha$ (all of which can be modified by gating through changes of refractive index, confinement, and excitonic response). The factor $\mathcal{P}_{\alpha}(V_g)$ encodes the \emph{phase} accumulated by the participating modes, which is sensitive to the graphene-induced changes in dispersion and phase matching.

For a waveguide or ring resonator section of length $L_{\alpha}$, the usual phase-matching factor for channel $\alpha$ is
\begin{equation}
\Phi_{\alpha}(V_g)
= \mathrm{sinc}\!\left[\frac{\Delta k_{\alpha}(V_g)\,L_{\alpha}}{2}\right]
e^{\,i\Delta k_{\alpha}(V_g)\,L_{\alpha}/2},
\end{equation}
where $\Delta k_{\alpha}(V_g)
= k_p(V_g) - \sum_{j=1}^n k_{j,\alpha}(V_g)$,
so that we can identify
\begin{equation}
\mathcal{P}_{\alpha}(V_g)
= \Phi_{\alpha}(V_g)
\equiv
\left|\Phi_{\alpha}(V_g)\right|
e^{i\varphi_{\alpha}(V_g)}.
\end{equation}
Here $k_p(V_g)$ and $k_{j,\alpha}(V_g)$ are the graphene- and Nb$_3$Cl$_8$-modified propagation constants of the pump and the $n$ output modes. Graphene gating changes their real parts (dispersion) and imaginary parts (loss), thus tuning both the magnitude $|\Phi_{\alpha}|$ and the phase $\varphi_{\alpha}$.

Combining these ingredients, the full complex coupling becomes
\begin{equation}
\kappa_{\alpha}(V_g)
=
|\kappa_{\alpha}(V_g)|\,e^{i\theta_{\alpha}(V_g)},
\end{equation}
where $\theta_{\alpha}(V_g)
=
\arg \chi^{(n)}_{\text{eff}}(V_g)
+ \varphi_{\alpha}(V_g)
+ \theta_{\alpha}^{(0)}$, and $\theta_{\alpha}^{(0)}$ is a static device-dependent phase (e.g.\ from fabrication). The key point is that \emph{both} the amplitude $|\kappa_{\alpha}(V_g)|$ and the phase $\theta_{\alpha}(V_g)$ can be tuned continuously by $V_g$, because (i) the complex susceptibility of Nb$_3$Cl$_8$ near excitonic resonances is modified by graphene screening, and (ii) the effective indices $n_{\text{eff}}(V_g)$ of the guided modes are shifted by the gate-controlled graphene conductivity. 

For GHZ-type generation [Sec.~\ref{subsec:GHZ}], we require two emission channels $\alpha=0,1$ with
\begin{equation}
\kappa_0(V_{g,0})
= |\kappa| e^{i\theta_0},
\qquad
\kappa_1(V_{g,1})
= |\kappa| e^{i\theta_1},
\end{equation}
so that the one-emission component of the output state becomes
\begin{equation}
|\psi_{\text{GHZ}}^{(1)}\rangle
\propto
|\kappa|\left(
\hat{A}_0^\dagger|0\rangle
+ e^{i\phi}\hat{A}_1^\dagger|0\rangle
\right),
\qquad
\phi = \theta_1 - \theta_0.
\end{equation}
By placing graphene contacts over two spatially separated sets of arms (each predominantly contributing to one of the channels) and biasing them independently at voltages $V_{g,0}$ and $V_{g,1}$, one can adjust $|\kappa_0|$ and $|\kappa_1|$ to be equal and then sweep the relative phase $\phi$ over (almost) the full range $[0,2\pi)$.

For more complex entangled states, such as cluster or general graph states [Sec.~\ref{subsec:cluster_states}], multiple emission channels $\alpha\in\mathcal{E}$ contribute with couplings $\kappa_{\alpha}(V_{g,\alpha})$. Tuning the gate voltages $V_{g,\alpha}$ region by region provides a way to program the effective \emph{weights} and \emph{phases} of the hyperedges in the interaction Hamiltonian,
\begin{equation}
\hat{H}_{\text{PDC}}^{(\text{graph})}
= i\hbar\sum_{\alpha\in\mathcal{E}}
\left[\kappa_{\alpha}(V_{g,\alpha})\,
\hat{A}_{\alpha}^\dagger
- \kappa_{\alpha}^*(V_{g,\alpha})\,\hat{A}_{\alpha}\right].
\end{equation}
After mapping modes to logical qubits via fixed linear optics, the resulting output state approximates a graph state whose adjacency matrix elements are proportional to $\kappa_{\alpha}(V_{g,\alpha})$. By varying the gate voltages, one can reconfigure the effective graph (turning edges ``on'' or ``off'' by changing $|\kappa_{\alpha}|$, and modifying their phases via $\theta_{\alpha}$), thereby enabling \emph{programmable} cluster-state generation from a single Nb$_3$Cl$_8$ device.

A further degree of freedom arises if the graphene gates are driven with time-dependent voltages $V_g(t)$ on a timescale comparable to or faster than the pump pulse separation. In that case, different pump pulses see different effective couplings $\kappa_{\alpha}[V_g(t_m)]$ at emission times $t_m$, producing a sequence of $n$-photon states with programmable phases and weights. Interpreting each emission time bin as a node in a temporal graph then leads to dynamically reconfigurable time-bin cluster states, which are particularly attractive for high-rate quantum communication.

\subsection{Summary of control knobs}
\label{subsec:summary_control}

To summarize, controlling the structure of the multiphoton entangled state generated by $n$-th order  in Nb$_3$Cl$_8$ relies on four main sets of knobs:

\begin{itemize}
  \item \textbf{Pump control:} amplitude, phase, polarization, spatial profile, and temporal structure (e.g.\ single vs multiple coherent pump pulses) determine which nonlinear tensor components and which sets of arms and modes are driven coherently.
  \item \textbf{Phase matching and dispersion engineering:} the 1$\times N$ splitter and waveguide geometry, together with the Nb$_3$Cl$_8$ layer, select which combinations of output modes are resonant and phase matched, thereby defining the allowed emission channels $\hat{A}_{\alpha}^\dagger$.
  \item \textbf{Linear optics:} on-chip interferometers implement unitaries $U$ that map the natural emission modes (arms of the splitter) to logical qubits and shape the interference between emission pathways, enabling GHZ, $W$, Dicke, and graph-state structures.
  \item \textbf{Graphene gating and heralding:} electrostatic gates on each arm tune the complex couplings $\kappa_{\alpha}$ (magnitudes and phases), while additional detectors and conditional logic project the multi-mode output onto desired subspaces, distilling specific entangled states from the raw emission pattern.
\end{itemize}

Because the ultra-large $\chi^{(n)}$ of monolayer Nb$_3$Cl$_8$ enhances the underlying $n$-photon  rate, these control techniques can be applied in realistic integrated photonic devices, opening a path toward on-chip generation of a broad family of multiphoton entangled resource states for quantum technologies.

\section{Projected device performance with 1$\times N$ splitters}
\label{sec:device_performance}

We now discuss how the quantum theory of Sec.~\ref{sec:quantum_theory_PDC} and the 1$\times N$ architecture of Fig.~\ref{fig:1x5_beamsplitter} translate into projected device performance when monolayer Nb$_3$Cl$_8$ is used as the active nonlinear medium. Rather than focusing on microrings or photonic-crystal cavities, we concentrate on straight-waveguide 1$\times N$ splitters and use the ab-initio susceptibilities of Sec.~\ref{sec:chi_abinitio} to estimate the relative performance of representative devices for three-photon GHZ states ($n=3$) and four-photon cluster states ($n=4$).

In all cases, the $n$-photon generation rate in a single arm or channel $\alpha$ is governed by Eq.~\eqref{eq:Rn_full},
\begin{equation}
R_n^{(\alpha)}
= 2\pi\,\frac{\hbar^n\,\epsilon_0^{2-n}}{2^n (n!)^2}\,
|\chi^{(n)}_{\text{eff}}|^2\,|E_p^{(\alpha)}|^2\,
\rho_{n,\alpha}^{(\text{eff})}\,
\prod_{j=1}^n
\frac{\omega_j}{n_j^2 V_{j,\alpha}},
\label{eq:Rn_alpha}
\end{equation}
where $E_p^{(\alpha)}$ is the pump field amplitude in arm $\alpha$, $V_{j,\alpha}$ the effective mode volumes, and $\rho_{n,\alpha}^{(\text{eff})}$ the effective $n$-photon density of states for that arm. The total rate is the coherent sum over all arms and channels included in the entangled state, as discussed in Sec.~\ref{sec:entangled_states}.

\subsection{Three-photon GHZ$_3$ generation with $n=3$}
\label{subsec:GHZ3_performance}

As a first example, consider the generation of a path-encoded three-photon GHZ state, using two arms of a 1$\times 3$ splitter (or two arms selected from a 1$\times 5$ device as in Fig.~\ref{fig:1x5_beamsplitter}). Third-order 3SPDC ($n=3$) in each arm produces a photon triplet at frequencies $(\omega_1,\omega_2,\omega_3)$ with complex amplitudes $\kappa_0$ and $\kappa_1$ in the two arms, which are then mapped to logical $|0\rangle_j$ and $|1\rangle_j$ modes by a fixed interferometer. As shown in Sec.~\ref{subsec:GHZ}, the one-emission component of the output state is
\begin{equation}
|\psi_{\text{GHZ}_3}^{(1)}\rangle
\propto
\kappa_0 |0,0,0\rangle
+ \kappa_1 |1,1,1\rangle.
\end{equation}
Tuning the gate voltages on the graphene contacts such that $|\kappa_0| = |\kappa_1| \equiv |\kappa|$ and controlling the relative phase $\phi = \arg(\kappa_1/\kappa_0)$ yields a GHZ$_3$ state \((|0,0,0\rangle + e^{i\phi}|1,1,1\rangle)/\sqrt{2}\).

The relevant nonlinear coefficient for three-photon 3SPDC is the third-order susceptibility. From Table~\ref{tab:chi} we have, for monolayer Nb$_3$Cl$_8$,
\begin{equation}
\chi^{(3)}_{\text{Nb}_3\text{Cl}_8} \simeq 3.4\times 10^{-18}~\mathrm{(m/V)^2}
\quad\text{at } E \approx 0.3~\mathrm{eV},
\end{equation}
while typical values for bulk silica (used in fiber-based 3SPDC) are $\chi^{(3)}_{\text{SiO}_2} \sim 2$--$3\times 10^{-22}$~(m/V)$^2$.\cite{Boyd_NLO_review} The ratio
\begin{equation}
\frac{|\chi^{(3)}_{\text{Nb}_3\text{Cl}_8}|}{|\chi^{(3)}_{\text{SiO}_2}|}
\sim 10^{4},
\end{equation}
translates into an enhancement factor
\begin{align}
\frac{R_3^{(\text{Nb}_3\text{Cl}_8)}}{R_3^{(\text{SiO}_2)}}
&\sim
\left|\frac{\chi^{(3)}_{\text{Nb}_3\text{Cl}_8}}
{\chi^{(3)}_{\text{SiO}_2}}\right|^2
\times
\frac{|E_p^{(\text{Nb}_3\text{Cl}_8)}|^2}{|E_p^{(\text{SiO}_2)}|^2}
\times
\frac{\rho_{3,\text{Nb}_3\text{Cl}_8}^{(\text{eff})}}{\rho_{3,\text{SiO}_2}^{(\text{eff})}}
\nn\\
&\times
\frac{\prod_j \omega_j/(n_j^2 V_{j,\text{Nb}_3\text{Cl}_8})}
{\prod_j \omega_j/(n_j^2 V_{j,\text{SiO}_2})}.
\label{eq:R3_enhancement}
\end{align}
For comparable pump intensities and overall mode volumes, the leading factor is $|\chi^{(3)}|^2$, giving a \emph{baseline} enhancement of order $10^8$ in the intrinsic three-photon coupling strength relative to silica fiber 3SPDC. Additional factors arise from tighter confinement in integrated waveguides (smaller $V_j$) and increased effective density of states when multiple arms and modes are involved, further boosting $R_3$.

In practice, the single-triplet generation probability per pump pulse,
\begin{equation}
P_3 \approx \sum_{\alpha=0,1} R_3^{(\alpha)}\,\tau_{\text{int}},
\end{equation}
must remain $\ll 1$ to stay in the low-gain regime, where $\tau_{\text{int}}$ is the effective interaction time (set by device length and group velocity). Equation~\eqref{eq:R3_enhancement} implies that, for a fixed low-gain $P_3$ (e.g.\ $10^{-4}$--$10^{-2}$ per pulse), Nb$_3$Cl$_8$ allows the same triplet probability to be reached with substantially shorter devices or lower pump powers than in conventional 3SPDC fibers. Conversely, for a given device footprint and pump power, Nb$_3$Cl$_8$-based 1$\times N$ splitters can in principle achieve triplet rates several orders of magnitude larger than silica-fiber-based triplet sources, which is highly advantageous for GHZ$_3$-based protocols.

\subsection{Four-photon generation for linear cluster states with $n=4$}
\label{subsec:cluster4_performance}

As a second example, consider the generation of four-photon entangled states such as a four-qubit GHZ$_4$ state or a linear four-qubit cluster state using a 1$\times 4$ splitter. In this case, each arm hosts a fourth-order 4SPDC process ($n=4$),
\begin{equation}
\omega_p \to \omega_1 + \omega_2 + \omega_3 + \omega_4,
\end{equation}
mediated by the fourth-order susceptibility $\chi^{(4)}$. From Table~\ref{tab:chi}, monolayer Nb$_3$Cl$_8$ exhibits
\begin{equation}
\chi^{(4)}_{\text{Nb}_3\text{Cl}_8} \simeq 2.2\times 10^{-24}~\mathrm{(m/V)^3}
\quad\text{at } E \approx 0.3~\mathrm{eV},
\end{equation}
while reported values for prototypical 2D semiconductors such as monolayer MoS$_2$ are around
\begin{equation}
\chi^{(4)}_{\text{MoS}_2} \simeq 1.1\times 10^{-27}~\mathrm{(m/V)^3},
\end{equation}
and estimates for bulk nonlinear crystals (LiNbO$_3$, AlGaAs, InP) are even smaller.\cite{Boyd_NLO_review,MoS2_SHG_THz,LiNbO3_microresonator} Thus,
\begin{equation}
\frac{|\chi^{(4)}_{\text{Nb}_3\text{Cl}_8}|}{|\chi^{(4)}_{\text{MoS}_2}|}
\sim 2\times 10^{3},
\end{equation}
implying an enhancement
\begin{equation}
\frac{R_4^{(\text{Nb}_3\text{Cl}_8)}}{R_4^{(\text{MoS}_2)}}
\sim
\left|\frac{\chi^{(4)}_{\text{Nb}_3\text{Cl}_8}}
{\chi^{(4)}_{\text{MoS}_2}}\right|^2
\sim 4\times 10^{6},
\end{equation}
for similar geometries and pump intensities. This shows that Nb$_3$Cl$_8$ is particularly advantageous for \emph{genuine} fourth-order 4SPDC processes, where conventional materials are effectively dark.

In the 1$\times 4$ architecture, four arms (A,B,C,D) each host a fourth-order 4SPDC process with complex couplings $\kappa_A,\kappa_B,\kappa_C,\kappa_D$ tuned by graphene gates, as described in Sec.~\ref{subsec:graphene_tunable}. The one-emission component of the state is
\begin{equation}
|\psi^{(1)}\rangle
\propto
\kappa_A \hat{B}_A^\dagger|0\rangle
+ \kappa_B \hat{B}_B^\dagger|0\rangle
+ \kappa_C \hat{B}_C^\dagger|0\rangle
+ \kappa_D \hat{B}_D^\dagger|0\rangle,
\end{equation}
where $\hat{B}_j^\dagger$ creates a four-photon Fock state in arm $j$. A subsequent 4$\times 4$ interferometer implements a unitary $U$ that maps $(\hat{B}_A^\dagger,\hat{B}_B^\dagger,\hat{B}_C^\dagger,\hat{B}_D^\dagger)$ to logical qubit modes, enabling two representative target states:

\begin{itemize}
  \item \emph{GHZ$_4$ state:} by suppressing two arms (e.g.\ C and D) and tuning $V_{g,A}$ and $V_{g,B}$ such that $|\kappa_A|=|\kappa_B|$ and $\phi=\arg(\kappa_B/\kappa_A)$ is adjustable, one obtains
  \begin{equation}
  |\psi^{(1)}\rangle
  \propto
  |0,0,0,0\rangle + e^{i\phi}|1,1,1,1\rangle,
  \end{equation}
  a four-photon GHZ state generated by a \emph{single} fourth-order 4SPDC event.
  \item \emph{Linear four-qubit cluster state:} by choosing $U$ so that different arms effectively implement nearest-neighbor CZ-like couplings between qubits, and by postselecting on specific detection patterns, one can project the four-photon component onto a linear cluster state
  \begin{equation}
  |C_4\rangle =
  \prod_{j=1}^3 \hat{U}_{\text{CZ}}^{(j,j+1)}
  |+\rangle_1|+\rangle_2|+\rangle_3|+\rangle_4,
  \end{equation}
  with the effective edge weights and phases set by $(\kappa_A,\kappa_B,\kappa_C,\kappa_D)$ and thus by the gate voltages $\{V_{g,j}\}$.
\end{itemize}

The extremely large $|\chi^{(4)}_{\text{Nb}_3\text{Cl}_8}|$ means that, for a fixed device length and pump power, the probability of a \emph{single} fourth-order 4SPDC event per pulse can be orders of magnitude larger than in MoS$_2$ or LiNbO$_3$ devices. Conversely, one can maintain the same low-gain probability as in conventional platforms while substantially reducing the device footprint or pump power, which is appealing for dense integration and thermal management.

\subsection{Implications for scalable architectures}
\label{subsec:scaling_performance}

The qualitative message of Eqs.~\eqref{eq:R3_enhancement} and \eqref{eq:Rn_alpha} is that the $n$-photon generation rate in Nb$_3$Cl$_8$-based 1$\times N$ devices scales as
\begin{equation}
R_n \propto |\chi^{(n)}_{\text{eff}}|^2\,|E_p|^2\, \left(\prod_j \frac{\omega_j}{V_j}\right)\,\rho_n^{(\text{eff})},
\end{equation}
with $|\chi^{(n)}_{\text{eff}}|^2$ providing an enormous lever arm for $n\ge 4$. Because the mode volumes $V_j$ in integrated waveguides are typically of order a few $\mu$m$^3$ and the effective density of states $\rho_n^{(\text{eff})}$ can be enhanced by using multiple arms and carefully engineered dispersion, Nb$_3$Cl$_8$-based 1$\times N$ splitters offer a realistic route to multi-photon sources in which genuine $n$-th order $n$SPDC (rather than cascaded pairwise processes) plays the central role.

When combined with the tunability afforded by graphene gates and reconfigurable interferometers, as described in Sec.~\ref{sec:entangled_states}, this enhanced nonlinear response enables electrically programmable, on-chip sources of GHZ, $W$, Dicke, and cluster states. While a full optimization of geometry, dispersion, and loss will be required for quantitative device design, the ab-initio susceptibilities of Sec.~\ref{sec:chi_abinitio} already indicate that Nb$_3$Cl$_8$ can transform high-order $n$SPDC from a fundamentally weak process into a practical resource for integrated quantum photonics.

\begin{figure}
	\begin{center}
		\includegraphics[width=3.4in]{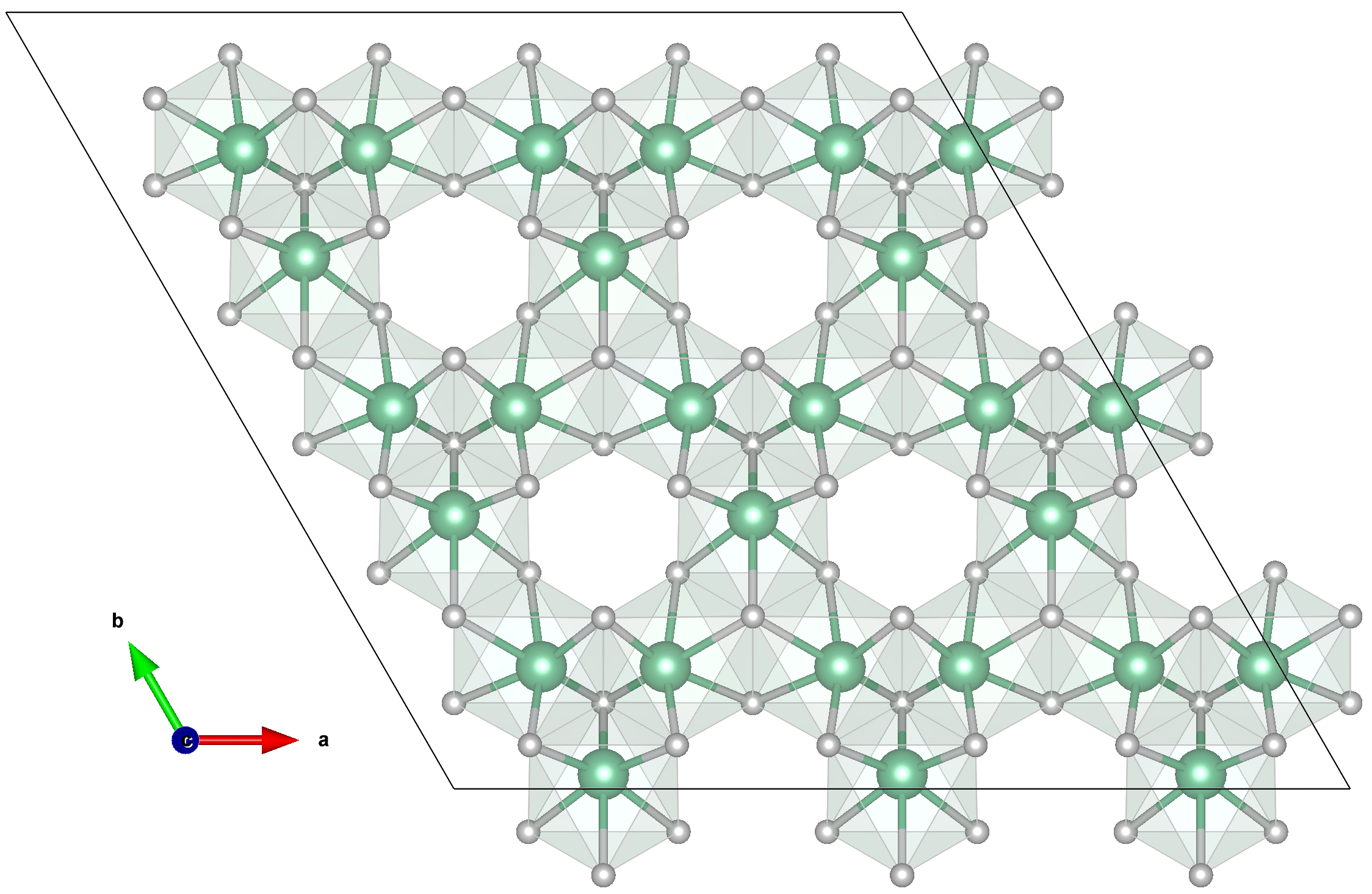}
	\end{center}
	\caption{Top view of 2D Nb$_3$Cl$_8$ monolayer. Nb atoms are green and Cl atoms are grey balls.}
	\label{fig:Nb3Cl8_ML}
\end{figure}

\section{Methods}

\subsection{Electronic-structure and exciton calculations}

The ground-state electronic structure of monolayer Nb$_3$Cl$_8$ is obtained within DFT using the PBE exchange–correlation functional and fully relativistic optimized norm-conserving Vanderbilt (ONCV) pseudopotentials including spin–orbit coupling, as implemented in \textsc{Quantum~Espresso}.\cite{QE1,QE2} A plane-wave cutoff of 450~Ry is used. The monolayer is modeled in a slab geometry with a vacuum spacing of 20~\AA\ and a truncated Coulomb interaction in the out-of-plane direction to avoid spurious interlayer screening. Structural parameters are relaxed until Hellmann–Feynman forces are below 0.001~eV/\AA.

Starting from the DFT wavefunctions, we perform eigenvalue-only self-consistent GW (evGW) calculations to obtain quasiparticle band structures, using the \textsc{Yambo} code.\cite{Sangalli2019} The screened Coulomb interaction is evaluated within the random-phase approximation, and convergence with respect to the number of empty bands and the frequency grid is carefully checked. The resulting quasiparticle energies are then used to construct the BSE Hamiltonian, which is solved on a $9\times 9\times 1$ $\mathbf{k}$-mesh including 12 valence and 12 conduction bands, yielding exciton energies and eigenvectors. The macroscopic dielectric function and absorption spectra are obtained from the exciton eigenstates and oscillator strengths. The random integration method is employed to properly converge the small-$q$ Coulomb singularity in the 2D geometry.

The linear optical response is obtained from the BSE, yielding the absorption spectrum and exciton oscillator strengths for in-plane and out-of-plane polarizations. The lowest dark Frenkel exciton appears as a strong feature at negative excitation energy in the exciton spectrum, while bright excitons with energies around 2~eV dominate the absorption.\cite{Khan_multiferroic_Nb3Cl8,Skachkov_GW_BSE_NL} These excitonic states provide the starting point for the time-dependent BSE / Kadanoff–Baym computations used to extract the high-order susceptibilities in the next subsection.

\begin{figure}
	\begin{center}
		\includegraphics[width=3.4in]{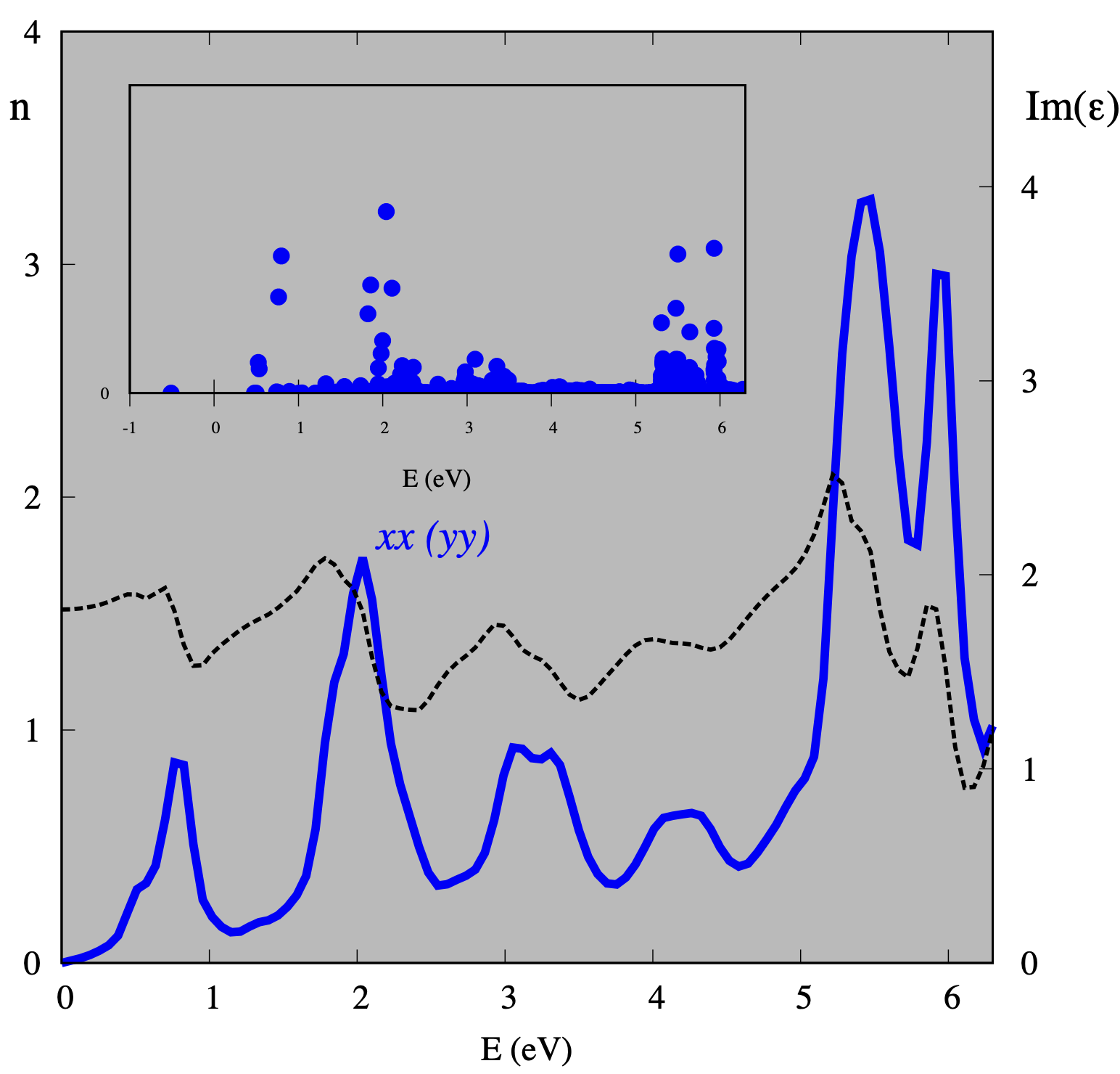}
	\end{center}
	\caption{The absorption spectra of ML Nb$_3$Cl$_8$, imaginary part of the dielectric function Im($\epsilon^{(1)}_{xx(yy)}$) (blue solid line), and refractive index n (dotted line). The insert: the oscillator strength as a function of energy.}
	\label{fig:LRx}
\end{figure}

\subsection{Time-dependent BSE and extraction of $\chi^{(n)}$}

To compute the nonlinear susceptibilities, we propagate the one-particle density matrix under a time-dependent electric field within the Kadanoff–Baym formalism, using the TD-BSE implementation in the \textsc{Yambo} code.\cite{Sangalli2019} The time evolution follows
\begin{equation}
i\hbar \frac{d}{dt} \rho_{nm\mathbf{k}}(t) = \left[H^{\text{QP}}_{\mathbf{k}} + H^{\text{HXC}}[\rho(t)] + H^{\text{ext}}(t), \rho(t)\right]_{nm},
\end{equation}
where \(H^{\text{QP}}_{\mathbf{k}}\) is the quasiparticle Hamiltonian, \(H^{\text{HXC}}\) is the Hartree–exchange–correlation contribution including screened electron–hole interactions in the BSE kernel, and \(H^{\text{ext}}(t) = -e \mathbf{E}(t)\cdot \mathbf{r}\) describes the coupling to the external electric field.

\begin{figure}
	\begin{center}
		\includegraphics[width=3.4in]{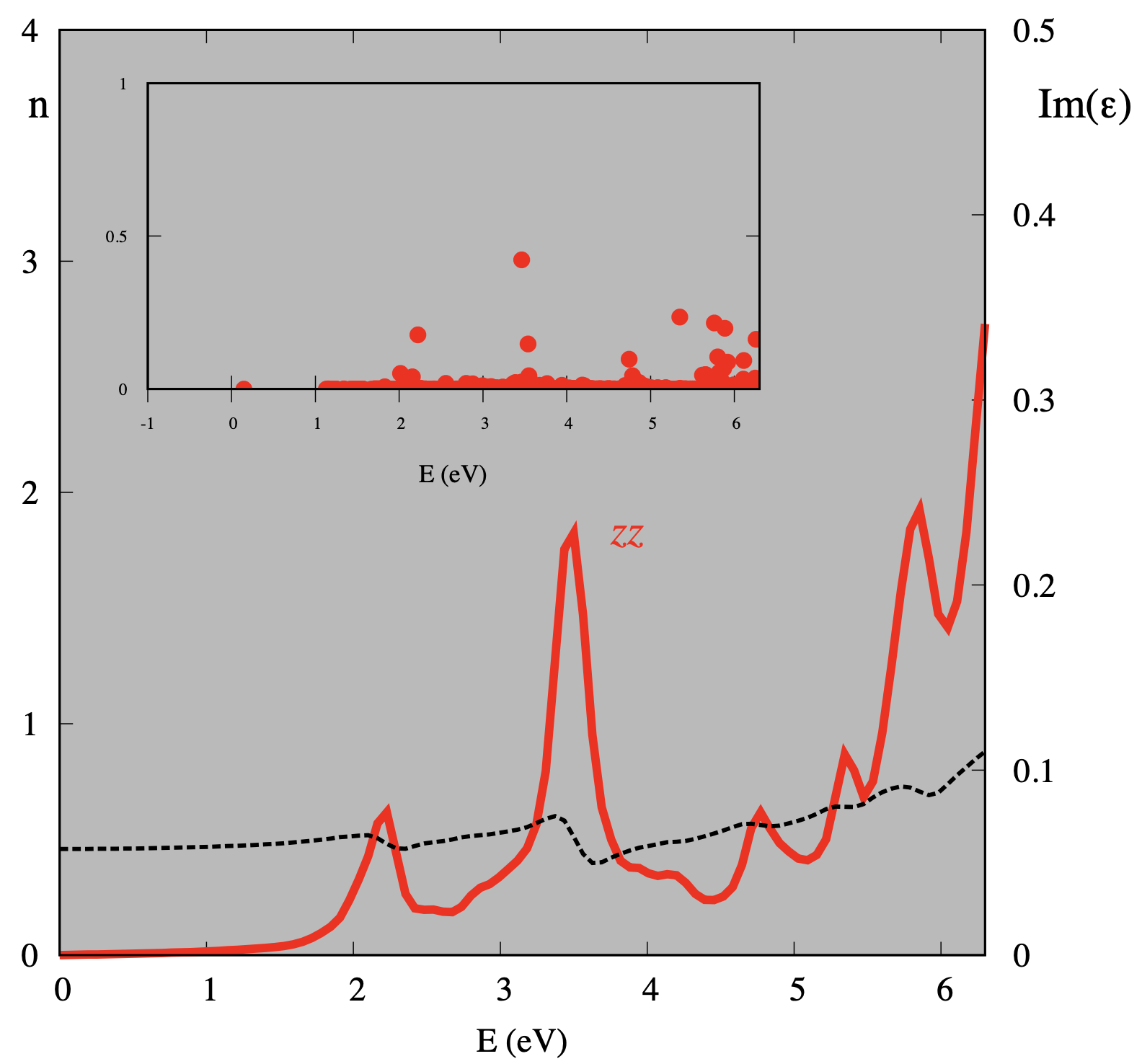}
	\end{center}
	\caption{The absorption spectra of ML Nb$_3$Cl$_8$, imaginary part of the dielectric function Im($\epsilon^{(1)}_{zz}$) (red solid line), and refractive index n (dotted line). The inset shows the oscillator strength as a function of energy.}
	\label{fig:LRz}
\end{figure}  

We consider two excitation protocols: (i) a continuous-wave sinusoidal field \(\mathbf{E}(t) = E_0 \hat{\mathbf{y}}\sin(\omega t)\) switched on at \(t=0\), and (ii) a combination of a short \(\delta\)-like pulse at \(t=0\) followed by the sinusoidal field starting at \(t=10\)~fs, which prepares the system in the dark Frenkel exciton state before probing its nonlinear response. The macroscopic polarization is computed from the density matrix via the Berry-connection formulation of the position operator in \(\mathbf{k}\)-space,
\begin{equation}
\mathbf{P}(t) = -\frac{e}{(2\pi)^2}\sum_{nm}\int_{\text{BZ}} d^2k\, \rho_{mn\mathbf{k}}(t)\, \mathbf{A}_{nm}(\mathbf{k}),
\end{equation}
where \(\mathbf{A}_{nm}(\mathbf{k}) = \langle u_{n\mathbf{k}}|i\nabla_{\mathbf{k}}|u_{m\mathbf{k}}\rangle\) is the Berry connection between Bloch states.

The Fourier transform \(\mathbf{P}(\omega)\) is analyzed to extract the harmonic components at frequencies \(m\omega\) (with \(m\le 7\)), and the corresponding susceptibilities \(\chi^{(m)}(\omega)\) are obtained by fitting the polarization response as a polynomial in the driving field amplitude:
\begin{equation}
P_\alpha(\omega) = \sum_{n} \epsilon_0 \chi^{(n)}_{\alpha\beta_1\ldots\beta_n}(\omega;\omega_1,\ldots,\omega_n)
\prod_{j=1}^{n} E_{\beta_j}(\omega_j),
\end{equation}
including appropriate symmetrization over permutations of the input frequencies \(\omega_j\). By comparing results for different field amplitudes and excitation protocols, we confirm the convergence of \(\chi^{(n)}\) up to \(n=7\) and identify the role of the excitonic ground state in enhancing the nonlinear response.

\section{Numerical Results}
\label{sec:numerical} 

\begin{figure}
\begin{center}
		\includegraphics[width=2in]{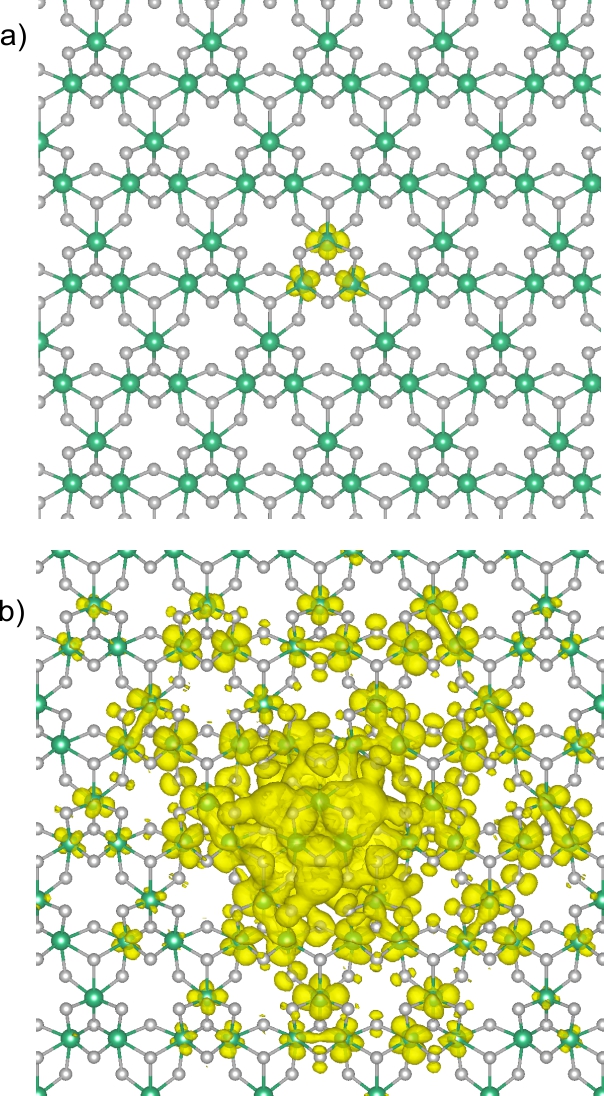}
	\end{center}
	\caption{Spatial distributions of (a) a localized dark Frenkel exciton at -0.51 eV and (b) a distributed bright exciton at 2.04 eV.}
	\label{fig:FE}
\end{figure}  

\begin{figure*}
\begin{center}
		\includegraphics[width=6in]{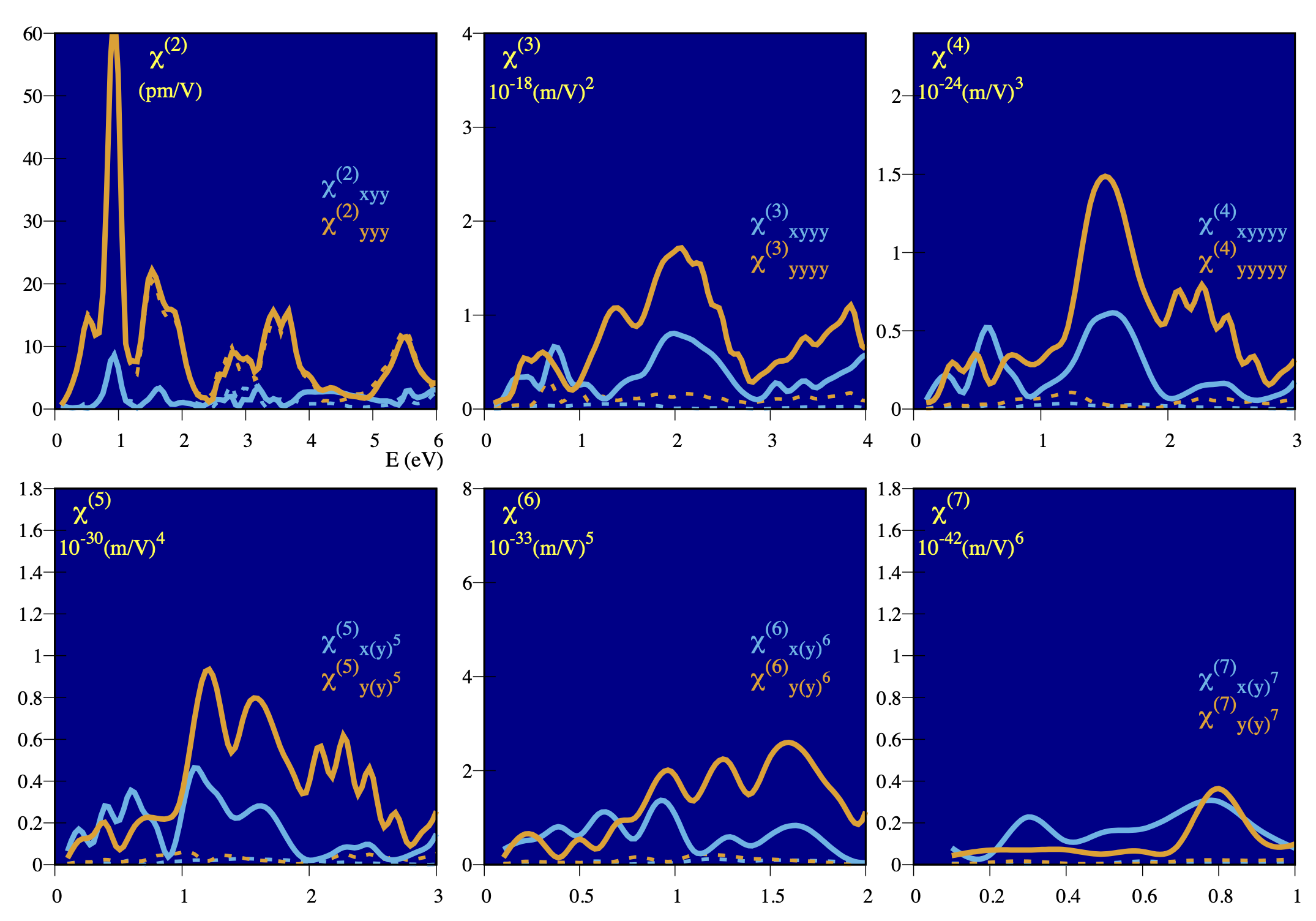}
	\end{center}
	\caption{Nonlinear susceptibilities $\chi^{(2)}$ - $\chi^{(7)}$ for sinusoidal perturbation (dashed) and delta pulse plus sinusoidal wave perturbations (solid lines).}
	\label{fig:NLR}
\end{figure*}

The linear response spectra of monolayer (ML) Nb$_3$Cl$_8$ obtained with BSE is shown on Fig.~\ref{fig:LRx} and \ref{fig:LRz} for electromagnetic wave perturbations in $x(y)$ and $z$ directions, respectively. The inserts on the Fig.~\ref{fig:LRx} and \ref{fig:LRz} show the oscillator strength of the excitations as a function of energy. The first dark exciton in Fig.~\ref{fig:LRx} has negative energy $-0.51$ eV. Fig.~\ref{fig:FE}a shows spatial distribution of this excitonic state, which shows strong localization of the dark Frenkel exciton. Fig.~\ref{fig:FE}b demonstrates spatial distribution of excitonic wave function for brightest exciton at 2.04 eV. 

Nonlinear optical response from ML $\mathrm{Nb}_3\mathrm{Cl}_8$  obtained with TD-BSE is represented in Fig.~\ref{fig:NLR}. We considered two different sets of perturbations. First, pure sinusoidal electromagnetic wave is switched on at $t=0$. For second perturbation, we first act on the system with a delta pulse at $t=0$, and then at $t=10 fs$ the sin wave perturbation is switching on. First delta pulse excites all possible frequencies in the system, and stimulate the system to go to lower energy dark excitonic state corresponding to $-0.51$ eV shown in the inset of Fig.~\ref{fig:LRx}. The following sin perturbation produces the NLR from the system which is in the ground excitonic state. The perturbing electromagnetic field is oriented along the $y$ direction, and the optical response of the electronic system is shown by the orange curves for the response in the $y$ direction and by the blue curves for the response in the $x$ direction. Dashed lines correspond to excitation from ground electronic state, whereas the solid lines correspond to excitation of the system from the ground excitonic state. Comparison of SHG and HHG with experimental values for MoS$_2$ shows that the third-order nonlinear response is comparable with MoS$_2$ for which it was reported the value of $3.8\times 10^{-18}$ (m/V)$^2$. \cite{Wang2022}  However, experimental values for fourth and fifth order susceptibilities for MoS$_2$, $\chi^{4}$ = $1.1\times 10^{-27}$ (m/V)$^3$ and $\chi^{5}$ = $6.6\times 10^{-36}$ (m/V)$^4$, are five and nine orders of magnitude smaller than for $\mathrm{Nb}_3\mathrm{Cl}_8$. 

\section{Conclusions and outlook}

We have proposed monolayer Nb$_3$Cl$_8$ as a 2D excitonic Mott insulator platform for high-order parametric downconversion and the generation of entangled photon multiplets. Using many-body GW--BSE and time-dependent BSE / Kadanoff--Baym simulations, we have computed the nonlinear susceptibilities $\chi^{(2)}$–$\chi^{(7)}$ of Nb$_3$Cl$_8$ and shown that its high-order coefficients $\chi^{(4)}$ and $\chi^{(5)}$ exceed those of prototypical 2D materials and bulk nonlinear crystals by several orders of magnitude. We have traced this enhancement to the unique combination of flat electronic bands, strongly bound Frenkel excitons with permanent out-of-plane dipole moments, and ferroelectric alignment of these excitons in the noncentrosymmetric monolayer.

Building on this \emph{ab initio} characterization, we developed a quantum-optical description of $n$-th order spontaneous parametric downconversion in a 2D nonlinear sheet and derived explicit expressions for the $n$-photon generation rate in terms of the microscopic susceptibilities, pump field, mode volumes, and an effective $n$-photon density of states. We then combined this theory with experimentally demonstrated 1$\times N$ integrated power splitters with arbitrary power ratios to propose Nb$_3$Cl$_8$-based 1$\times N$ architectures in which each output arm hosts a monolayer Nb$_3$Cl$_8$ patch, optionally gated by graphene. In this configuration, the 1$\times N$ splitter defines the spatial mode basis, the giant $\chi^{(n)}$ of Nb$_3$Cl$_8$ sets the overall $n$-photon downconversion strength, and graphene gates provide independent electrical control of the complex $n$-photon amplitudes $\kappa_\alpha$ in each arm.

Within this framework, we showed how suitable interferometric networks and heralding can realize a broad family of multiphoton entangled states, including GHZ, $W$, Dicke, and cluster states. In particular, we analyzed three-photon GHZ$_3$ states generated by genuine third-order 3SPDC ($n=3$) in two arms of a 1$\times N$ splitter, and four-photon GHZ$_4$ and linear cluster states generated by fourth-order 4SPDC ($n=4$) in four arms. Using the computed $\chi^{(3)}$ and $\chi^{(4)}$ values for Nb$_3$Cl$_8$, we estimated that the intrinsic $n$-photon coupling strengths can exceed those of silica-fiber- and MoS$_2$-based devices by factors of $\sim 10^8$ and $\sim 10^6$, respectively, under comparable geometric conditions, indicating that truly high-order $n$SPDC processes (rather than cascaded pairwise processes) can become experimentally accessible in realistic integrated devices.

Beyond Nb$_3$Cl$_8$ itself, our results naturally extend to the broader family of breathing-kagome trimer halides. The niobium halides Nb$_3$X$_8$ (X = Cl, Br, I) share the same cluster lattice and exhibit similar flat-band and correlated physics.\cite{Haraguchi_JPSJ2024,Sun_flatbands,Feng_Nb3X8_triferroics} Angle-resolved photoemission spectroscopy has revealed flat and weakly dispersive bands in Nb$_3$Br$_8$ and Nb$_3$I$_8$,\cite{Regmi_Nb3Br8_flatbands,Regmi_Nb3I8_flatbands} while first-principles calculations predict that Nb$_3$X$_8$ monolayers are multiferroic semiconductors with coupled ferroelectric, ferromagnetic, and ferrovalley orders.\cite{Feng_Nb3X8_triferroics} Closely related tantalum halides Ta$_3$X$_8$ (X = Cl, Br, I) also form breathing kagome lattices and have been predicted to host triferroic phases, valley-polarized states, and even spin-polarized triplet excitonic insulators in Ta$_3$I$_8$ and Ta$_3$Br$_8$.\cite{Li_Ta3X8_multiferroics,Xing_Ta3I8_triferroic,Lu_Ta3I8_CBV,Sheng_Ta3X8_triplet_EI} Given the common structural motif and the strong excitonic and multiferroic responses in these families, it is reasonable to expect similarly large high-order nonlinear coefficients in Nb$_3$Br$_8$, Nb$_3$I$_8$, and Ta$_3$X$_8$ monolayers. We therefore anticipate that the programmable 1$\times N$ entangled multiphoton state generators proposed here can be implemented across a whole class of Nb and Ta halide breathing-kagome semiconductors, with the different halides providing additional spectral and spin–valley tuning knobs.

Future work will include (i) full-wave electromagnetic simulations of specific 1$\times N$ splitter geometries incorporating Nb$_3$Cl$_8$ and related halides, with optimized mode confinement and dispersion engineering for targeted $n$-photon states; (ii) quantitative assessment of loss, noise, and multi-pair emission in the high-order $n$SPDC regime; (iii) exploration of electrostatic, strain, and magnetic tuning of the excitonic resonances and nonlinear response, including dynamic gate control for time-bin cluster states; and (iv) experimental implementation of Nb$_3$Cl$_8$- and Nb$_3$X$_8$/Ta$_3$X$_8$-based multiphoton sources on existing silicon nitride and lithium niobate platforms, followed by characterization of their entanglement properties. More broadly, our results suggest that correlated flat-band materials with excitonic Mott insulating or excitonic-insulator phases, when combined with programmable integrated photonics, are fertile grounds for discovering ultra-strong nonlinearities and realizing electrically reconfigurable, high-order quantum nonlinear optical devices.

\section*{Data availability}

The data supporting the findings of this study, including GW–BSE and TD-BSE input files, convergence tests, and processed nonlinear susceptibility spectra, will be made available in a public repository \cite{GitHub} upon publication or can be obtained from the corresponding author upon reasonable request.

\section*{Acknowledgments}
M. N. L. and D. R. E. acknowledge support by the Air Force Office of Scientific Research (AFOSR) under award no. FA9550-23-1-0472.
D. S. and M. N. L. acknowledge support by the AFOSR under award no. FA9550-23-1-0455.
Calculations were performed at the Stokes high performance computer cluster of the University of Central Florida.
Some calculations were performed on the Darwin high performance computer cluster provided by the ACCESS program of the National Science Foundation (NSF).
M. N. L. and D.S. acknowledge support by the NSF ACCESS program under allocation no. PHY230182.

\clearpage
\section*{References}
\bibliography{bibliography}

\begin{thebibliography}{10}
\expandafter\ifx\csname url\endcsname\relax
  \def\url#1{\texttt{#1}}\fi
\expandafter\ifx\csname urlprefix\endcsname\relax\def\urlprefix{URL }\fi
\providecommand{\bibinfo}[2]{#2}
\providecommand{\eprint}[2][]{\url{#2}}

\bibitem{Centini2005}
\bibinfo{author}{Centini, M.} \emph{et~al.}
\newblock \bibinfo{title}{Entangled photon pair generation by spontaneous parametric down-conversion in finite-length one-dimensional photonic crystals}.
\newblock \emph{\bibinfo{journal}{Phys. Rev. A}} \textbf{\bibinfo{volume}{72}}, \bibinfo{pages}{033806} (\bibinfo{year}{2005}).

\bibitem{Couteau2018}
\bibinfo{author}{Couteau, C.}
\newblock \bibinfo{title}{Spontaneous parametric down-conversion}.
\newblock \emph{\bibinfo{journal}{Contemp. Phys.}} \textbf{\bibinfo{volume}{59}}, \bibinfo{pages}{291--304} (\bibinfo{year}{2018}).

\bibitem{Hutter2020}
\bibinfo{author}{Hutter, L.}, \bibinfo{author}{Lima, G.} \& \bibinfo{author}{Walborn, S.}
\newblock \bibinfo{title}{Boosting entanglement generation in down-conversion with incoherent illumination}.
\newblock \emph{\bibinfo{journal}{Phys. Rev. Lett.}} \textbf{\bibinfo{volume}{125}}, \bibinfo{pages}{193602} (\bibinfo{year}{2020}).

\bibitem{Zeng2007}
\bibinfo{author}{Zeng, Y.}, \bibinfo{author}{Fu, Y.}, \bibinfo{author}{Chen, X.}, \bibinfo{author}{Lu, W.} \& \bibinfo{author}{Ågren, H.}
\newblock \bibinfo{title}{Highly efficient generation of entangled photon pair by spontaneous parametric downconversion in defective photonic crystals}.
\newblock \emph{\bibinfo{journal}{J. Amer. Optic. Soc. B}} \textbf{\bibinfo{volume}{24}}, \bibinfo{pages}{1365--1368} (\bibinfo{year}{2007}).
\newblock \urlprefix\url{https://opg.optica.org/josab/abstract.cfm?URI=josab-24-6-1365}.

\bibitem{Liu2021}
\bibinfo{author}{Liu, Y.-C.} \emph{et~al.}
\newblock \bibinfo{title}{Observation of frequency‑uncorrelated photon pairs generated by counter‑propagating spontaneous parametric down‑conversion}.
\newblock \emph{\bibinfo{journal}{Sci. Reports}} \textbf{\bibinfo{volume}{11}}, \bibinfo{pages}{12628} (\bibinfo{year}{2021}).

\bibitem{Slattery2019}
\bibinfo{author}{Slattery, O.}, \bibinfo{author}{Ma, L.}, \bibinfo{author}{Zong, K.} \& \bibinfo{author}{Tang, X.}
\newblock \bibinfo{title}{Background and review of cavity-enhanced spontaneous parametric down-conversion}.
\newblock \emph{\bibinfo{journal}{J. Res. National Inst. Stand. Techn.}} \textbf{\bibinfo{volume}{124}}, \bibinfo{pages}{124019} (\bibinfo{year}{2019}).

\bibitem{Afsharnia2024}
\bibinfo{author}{Afsharnia, M.} \emph{et~al.}
\newblock \bibinfo{title}{Generation of infrared photon pairs by spontaneous four‑wave mixing in a {CS2}‑filled microstructured optical fiber}.
\newblock \emph{\bibinfo{journal}{Sci. Reports}} \textbf{\bibinfo{volume}{14}}, \bibinfo{pages}{977} (\bibinfo{year}{2024}).

\bibitem{Cui2012}
\bibinfo{author}{Cui, L.}, \bibinfo{author}{Li, X.} \& \bibinfo{author}{Zhao, N.}
\newblock \bibinfo{title}{Spectral properties of photon pairs generated by spontaneous four-wave mixing in inhomogeneous photonic crystal fibers}.
\newblock \emph{\bibinfo{journal}{Phys. Rev. A}} \textbf{\bibinfo{volume}{85}}, \bibinfo{pages}{023825} (\bibinfo{year}{2012}).

\bibitem{Wang2024_4}
\bibinfo{author}{Wang, H.}, \bibinfo{author}{Zeng, Q.}, \bibinfo{author}{Ma, H.} \& \bibinfo{author}{Yuan, Z.}
\newblock \bibinfo{title}{Progress on chip-based spontaneous four-wave mixing quantum light sources}.
\newblock \emph{\bibinfo{journal}{Adv. Devices \& Instrumentation}} \textbf{\bibinfo{volume}{5}}, \bibinfo{pages}{0032} (\bibinfo{year}{2024}).

\bibitem{Pirandola2020}
\bibinfo{author}{Pirandola, S.} \emph{et~al.}
\newblock \bibinfo{title}{Advances in quantum cryptography}.
\newblock \emph{\bibinfo{journal}{Adv. Opt. Photonics}} \textbf{\bibinfo{volume}{12}}, \bibinfo{pages}{1012} (\bibinfo{year}{2020}).

\bibitem{Zhang2025Q}
\bibinfo{author}{Zhang, H.} \emph{et~al.}
\newblock \bibinfo{title}{Towards global quantum key distribution}.
\newblock \emph{\bibinfo{journal}{Nat. Reviews Electrical Engineering}}  (\bibinfo{year}{2025}).

\bibitem{Xu2020}
\bibinfo{author}{Xu, F.}, \bibinfo{author}{X., M.}, \bibinfo{author}{Zhang, Q.}, \bibinfo{author}{Lo, H.-K.} \& \bibinfo{author}{Pan, J.-W.}
\newblock \bibinfo{title}{Secure quantum key distribution with realistic devices}.
\newblock \emph{\bibinfo{journal}{Rev. Modern Physics}} \textbf{\bibinfo{volume}{92}}, \bibinfo{pages}{025002} (\bibinfo{year}{2020}).

\bibitem{Brazaola2024}
\bibinfo{author}{Brazaola-Vicario, A.}, \bibinfo{author}{Ruiz, A.}, \bibinfo{author}{Lage, O.}, \bibinfo{author}{Jacob, E.} \& \bibinfo{author}{Astorga, J.}
\newblock \bibinfo{title}{Quantum key distribution: a survey on current vulnerability trends and potential implementation risks}.
\newblock \emph{\bibinfo{journal}{Optics Continuum}} \textbf{\bibinfo{volume}{3}}, \bibinfo{pages}{1438} (\bibinfo{year}{2024}).

\bibitem{Zapatero2023}
\bibinfo{author}{Zapatero, V.} \emph{et~al.}
\newblock \bibinfo{title}{Advances in device-independent quantum key distribution}.
\newblock \emph{\bibinfo{journal}{npj quantum information}} \textbf{\bibinfo{volume}{9}}, \bibinfo{pages}{10} (\bibinfo{year}{2023}).

\bibitem{Montenegro2025}
\bibinfo{author}{Montenegro, V.} \emph{et~al.}
\newblock \bibinfo{title}{Review: Quantum metrology and sensing with many-body systems}.
\newblock \emph{\bibinfo{journal}{Physics Reports}} \textbf{\bibinfo{volume}{1134}}, \bibinfo{pages}{1--62} (\bibinfo{year}{2025}).

\bibitem{Deng2024}
\bibinfo{author}{Deng, X.} \emph{et~al.}
\newblock \bibinfo{title}{Quantum-enhanced metrology with large {Fock} states}.
\newblock \emph{\bibinfo{journal}{Nature Physics}} \textbf{\bibinfo{volume}{20}}, \bibinfo{pages}{1874--1881} (\bibinfo{year}{2024}).

\bibitem{Huang2024}
\bibinfo{author}{Huang, J.}, \bibinfo{author}{Zhunag, M.} \& \bibinfo{author}{Lee, C.}
\newblock \bibinfo{title}{Entanglement-enhanced quantum metrology: From standard quantum limit to {Heisenberg} limit}.
\newblock \emph{\bibinfo{journal}{Appl. Phys. Reviews}} \textbf{\bibinfo{volume}{11}}, \bibinfo{pages}{031302} (\bibinfo{year}{2024}).

\bibitem{Slussarenko2019}
\bibinfo{author}{Slussarenko, S.} \& \bibinfo{author}{Pryde, G.}
\newblock \bibinfo{title}{Photonic quantum information processing: A concise review}.
\newblock \emph{\bibinfo{journal}{Appl. Phys. Reviews}} \textbf{\bibinfo{volume}{6}}, \bibinfo{pages}{041303} (\bibinfo{year}{2019}).

\bibitem{Flamini2019}
\bibinfo{author}{Flamini, F.}, \bibinfo{author}{Spagnolo, N.} \& \bibinfo{author}{Sciarrino, F.}
\newblock \bibinfo{title}{Photonic quantum information processing: a review}.
\newblock \emph{\bibinfo{journal}{Reports on Progress in Physics}} \textbf{\bibinfo{volume}{82}}, \bibinfo{pages}{016001} (\bibinfo{year}{2019}).

\bibitem{Delfanazari2025}
\bibinfo{author}{Delfanazari, K.}
\newblock \bibinfo{title}{Chip-scale electrically driven superconducting coherent photon sources for quantum information processing}.
\newblock \emph{\bibinfo{journal}{Nature Photonics}} \textbf{\bibinfo{volume}{19}}, \bibinfo{pages}{1163} (\bibinfo{year}{2025}).

\bibitem{Domininguez2020}
\bibinfo{author}{Domínguez-Serna, F.}, \bibinfo{author}{U’Ren, A.} \& \bibinfo{author}{Garay-Palmett, K.}
\newblock \bibinfo{title}{Third-order parametric down-conversion: A stimulated approach}.
\newblock \emph{\bibinfo{journal}{Phys. Rev. A}} \textbf{\bibinfo{volume}{101}}, \bibinfo{pages}{033813} (\bibinfo{year}{2020}).

\bibitem{Bacaoco_TOPDC}
\bibinfo{author}{Bacaoco, M.~Y.}, \bibinfo{author}{Koshelev, K.} \& \bibinfo{author}{Solntsev, A.~S.}
\newblock \bibinfo{title}{Third-order spontaneous parametric down conversion in dielectric nonlinear resonant metasurfaces}.
\newblock \emph{\bibinfo{journal}{ACS Photonics}} \textbf{\bibinfo{volume}{12}} (\bibinfo{year}{2025}).
\newblock \bibinfo{note}{ArXiv:2504.02267}.

\bibitem{Cavanna2020}
\bibinfo{author}{Cavanna, A.} \emph{et~al.}
\newblock \bibinfo{title}{Progress toward third-order parametric down-conversion in optical fibers}.
\newblock \emph{\bibinfo{journal}{Phys. Rev.A}} \textbf{\bibinfo{volume}{101}}, \bibinfo{pages}{033840} (\bibinfo{year}{2020}).

\bibitem{Corona_TOSPDC}
\bibinfo{author}{Corona, M.}, \bibinfo{author}{Garay-Palmett, K.} \& \bibinfo{author}{{U'Ren}, A.~B.}
\newblock \bibinfo{title}{Experimental proposal for the generation of entangled photon triplets by third-order spontaneous parametric downconversion in optical fibers}.
\newblock \emph{\bibinfo{journal}{Optics Letters}} \textbf{\bibinfo{volume}{36}}, \bibinfo{pages}{190--192} (\bibinfo{year}{2011}).

\bibitem{Bertrand2025}
\bibinfo{author}{Bertrand, J.}, \bibinfo{author}{Boutou, V.}, \bibinfo{author}{Felix, C.}, \bibinfo{author}{Jegouso, D.} \& \bibinfo{author}{Boulanger, B.}
\newblock \bibinfo{title}{Experimental demonstration and modeling of near-infrared nonlinear third-order triple-photon generation stimulated over one mode}.
\newblock \emph{\bibinfo{journal}{APL Quantum}} \textbf{\bibinfo{volume}{2}}, \bibinfo{pages}{026114} (\bibinfo{year}{2025}).

\bibitem{optical_fiber_TOSPDC}
\bibinfo{author}{Corona, M.}, \bibinfo{author}{Garay-Palmett, K.} \& \bibinfo{author}{{U'Ren}, A.~B.}
\newblock \bibinfo{title}{Third-order spontaneous parametric down-conversion in thin optical fibers as a photon-triplet source}.
\newblock \emph{\bibinfo{journal}{Physical Review A}} \textbf{\bibinfo{volume}{84}}, \bibinfo{pages}{033823} (\bibinfo{year}{2011}).

\bibitem{Chang_3photon_SPDC}
\bibinfo{author}{Chang, C. W.~S.} \emph{et~al.}
\newblock \bibinfo{title}{Observation of three-photon spontaneous parametric downconversion in a superconducting parametric cavity}.
\newblock \emph{\bibinfo{journal}{Physical Review X}} \textbf{\bibinfo{volume}{10}}, \bibinfo{pages}{011011} (\bibinfo{year}{2020}).

\bibitem{Bao2024}
\bibinfo{author}{Bao, Z.} \emph{et~al.}
\newblock \bibinfo{title}{Creating and controlling global {Greenberger}-{Horne}-{Zeilinger} entanglement on quantum processors}.
\newblock \emph{\bibinfo{journal}{Nat. Commun.}} \textbf{\bibinfo{volume}{15}}, \bibinfo{pages}{8823} (\bibinfo{year}{2024}).

\bibitem{Chen2024}
\bibinfo{author}{Chen, L.} \emph{et~al.}
\newblock \bibinfo{title}{Observation of quantum nonlocality in {Greenberger}-{Horne}-{Zeilinger} entanglement on a silicon chip}.
\newblock \emph{\bibinfo{journal}{Optics Express}} \textbf{\bibinfo{volume}{32}}, \bibinfo{pages}{14904} (\bibinfo{year}{2024}).

\bibitem{Cao2024}
\bibinfo{author}{Cao, H.} \emph{et~al.}
\newblock \bibinfo{title}{Photonic source of heralded {Greenberger}-{Horne}-{Zeilinger} states}.
\newblock \emph{\bibinfo{journal}{Phys. Rev. Lett.}} \textbf{\bibinfo{volume}{132}}, \bibinfo{pages}{130604} (\bibinfo{year}{2024}).

\bibitem{Dur2000_Wclass}
\bibinfo{author}{D{\"u}r, W.}, \bibinfo{author}{Vidal, G.} \& \bibinfo{author}{Cirac, J.~I.}
\newblock \bibinfo{title}{Three qubits can be entangled in two inequivalent ways}.
\newblock \emph{\bibinfo{journal}{Physical Review A}} \textbf{\bibinfo{volume}{62}}, \bibinfo{pages}{062314} (\bibinfo{year}{2000}).

\bibitem{Eibl2004_expW}
\bibinfo{author}{Eibl, M.}, \bibinfo{author}{Kiesel, N.}, \bibinfo{author}{Bourennane, M.}, \bibinfo{author}{Kurtsiefer, C.} \& \bibinfo{author}{Weinfurter, H.}
\newblock \bibinfo{title}{Experimental realization of a three-qubit entangled $w$ state}.
\newblock \emph{\bibinfo{journal}{Physical Review Letters}} \textbf{\bibinfo{volume}{92}}, \bibinfo{pages}{077901} (\bibinfo{year}{2004}).

\bibitem{Menotti2016_energyW}
\bibinfo{author}{Menotti, M.}, \bibinfo{author}{Maccone, L.}, \bibinfo{author}{Sipe, J.~E.} \& \bibinfo{author}{Liscidini, M.}
\newblock \bibinfo{title}{Generation of energy-entangled $w$ states via parametric fluorescence in integrated devices}.
\newblock \emph{\bibinfo{journal}{Physical Review A}} \textbf{\bibinfo{volume}{94}}, \bibinfo{pages}{022313} (\bibinfo{year}{2016}).

\bibitem{Dicke1954}
\bibinfo{author}{Dicke, R.~H.}
\newblock \bibinfo{title}{Coherence in spontaneous radiation processes}.
\newblock \emph{\bibinfo{journal}{Physical Review}} \textbf{\bibinfo{volume}{93}}, \bibinfo{pages}{99--110} (\bibinfo{year}{1954}).

\bibitem{Stockton2003_Dicke}
\bibinfo{author}{Stockton, J.~K.}, \bibinfo{author}{Geremia, J.~M.}, \bibinfo{author}{Doherty, A.~C.} \& \bibinfo{author}{Mabuchi, H.}
\newblock \bibinfo{title}{Characterizing the entanglement of symmetric many-particle spin-$1/2$ systems}.
\newblock \emph{\bibinfo{journal}{Physical Review A}} \textbf{\bibinfo{volume}{67}}, \bibinfo{pages}{022112} (\bibinfo{year}{2003}).

\bibitem{Toth2007_Dicke}
\bibinfo{author}{T{\'o}th, G.}
\newblock \bibinfo{title}{Detection of entanglement in optical lattices of spin-$1/2$ particles}.
\newblock \emph{\bibinfo{journal}{Journal of the Optical Society of America B}} \textbf{\bibinfo{volume}{24}}, \bibinfo{pages}{275--282} (\bibinfo{year}{2007}).

\bibitem{Walschaers2021}
\bibinfo{author}{Walschaers, M.}
\newblock \bibinfo{title}{Non-gaussian quantum states and where to find them}.
\newblock \emph{\bibinfo{journal}{PRX Quantum}} \textbf{\bibinfo{volume}{2}}, \bibinfo{pages}{030204} (\bibinfo{year}{2021}).

\bibitem{Kawasaki2024}
\bibinfo{author}{Kawasaki, A.} \emph{et~al.}
\newblock \bibinfo{title}{Broadband generation and tomography of non-gaussian states for ultra-fast optical quantum processors}.
\newblock \emph{\bibinfo{journal}{Nature Communications}} \textbf{\bibinfo{volume}{15}}, \bibinfo{pages}{9075} (\bibinfo{year}{2024}).

\bibitem{Nielsen2006}
\bibinfo{author}{Nielsen, M.}
\newblock \bibinfo{title}{Cluster-state quantum computation}.
\newblock \emph{\bibinfo{journal}{Reports on Mathematical Physics}} \textbf{\bibinfo{volume}{57}}, \bibinfo{pages}{147} (\bibinfo{year}{2006}).

\bibitem{OSullivan2025}
\bibinfo{author}{O’Sullivan, J.} \emph{et~al.}
\newblock \bibinfo{title}{Deterministic generation of two-dimensional multi-photon cluster states}.
\newblock \emph{\bibinfo{journal}{Nature Communications}} \textbf{\bibinfo{volume}{16}}, \bibinfo{pages}{5505} (\bibinfo{year}{2025}).

\bibitem{Freund2025}
\bibinfo{author}{Freund, J.}, \bibinfo{author}{Pirker, A.}, \bibinfo{author}{Vandré, L.} \& \bibinfo{author}{Dür, W.}
\newblock \bibinfo{title}{Graph state extraction from two-dimensional cluster states}.
\newblock \emph{\bibinfo{journal}{New J. Physics}} \textbf{\bibinfo{volume}{16}}, \bibinfo{pages}{5505} (\bibinfo{year}{2025}).

\bibitem{Lyubarov2025}
\bibinfo{author}{Lyubarov, M.} \emph{et~al.}
\newblock \bibinfo{title}{Continuous-variables cluster states in photonic time-crystals}.
\newblock \emph{\bibinfo{journal}{Optica Quantum}} \textbf{\bibinfo{volume}{3}}, \bibinfo{pages}{366} (\bibinfo{year}{2025}).

\bibitem{Shi2021}
\bibinfo{author}{Shi, Y.} \& \bibinfo{author}{Waks, E.}
\newblock \bibinfo{title}{Deterministic generation of multidimensional photonic cluster states using time-delay feedback}.
\newblock \emph{\bibinfo{journal}{Phys. Rev. A}} \textbf{\bibinfo{volume}{104}}, \bibinfo{pages}{013703} (\bibinfo{year}{2021}).

\bibitem{Xu2017_OL}
\bibinfo{author}{Xu, K.} \emph{et~al.}
\newblock \bibinfo{title}{Integrated photonic power divider with arbitrary power ratios}.
\newblock \emph{\bibinfo{journal}{Optics Letters}} \textbf{\bibinfo{volume}{42}}, \bibinfo{pages}{855--858} (\bibinfo{year}{2017}).

\bibitem{Xu2016_arxiv}
\bibinfo{author}{Xu, K.} \emph{et~al.}
\newblock \bibinfo{title}{All passive photonic power divider with arbitrary split ratio}.
\newblock \emph{\bibinfo{journal}{arXiv preprint}}  (\bibinfo{year}{2016}).
\newblock \eprint{1609.03823}.

\bibitem{Franz2021_JOpt}
\bibinfo{author}{Franz, Y.} \& \bibinfo{author}{Guasoni, M.}
\newblock \bibinfo{title}{Compact 1 $\times$ {N} power splitters with arbitrary power ratio for integrated multimode photonics}.
\newblock \emph{\bibinfo{journal}{Journal of Optics}} \textbf{\bibinfo{volume}{23}}, \bibinfo{pages}{095802} (\bibinfo{year}{2021}).

\bibitem{Ma2020_SciRep}
\bibinfo{author}{Ma, H.-S.}, \bibinfo{author}{Ma, Y.-C.}, \bibinfo{author}{Li, Z.-Y.} \emph{et~al.}
\newblock \bibinfo{title}{Inverse-designed arbitrary-input and ultra-compact 1 $\times$ {N} power splitters based on high symmetric structure}.
\newblock \emph{\bibinfo{journal}{Scientific Reports}} \textbf{\bibinfo{volume}{10}}, \bibinfo{pages}{11757} (\bibinfo{year}{2020}).
\newblock \bibinfo{note}{Article number 11757}.

\bibitem{Liu2025_Nanomaterials}
\bibinfo{author}{Liu, Y.}, \bibinfo{author}{Wu, J.}, \bibinfo{author}{Wang, Z.} \emph{et~al.}
\newblock \bibinfo{title}{Inverse design of multi-port power splitter with arbitrary power ratio}.
\newblock \emph{\bibinfo{journal}{Nanomaterials}} \textbf{\bibinfo{volume}{15}}, \bibinfo{pages}{393} (\bibinfo{year}{2025}).

\bibitem{Haines2024_Nanophotonics}
\bibinfo{author}{Haines, J.} \emph{et~al.}
\newblock \bibinfo{title}{Fabrication of 1 $\times$ {N} integrated power splitters with arbitrary power ratio for single and multimode photonics}.
\newblock \emph{\bibinfo{journal}{Nanophotonics}} \textbf{\bibinfo{volume}{13}}, \bibinfo{pages}{339--348} (\bibinfo{year}{2024}).

\bibitem{Sun_flatbands}
\bibinfo{author}{Sun, Z.} \emph{et~al.}
\newblock \bibinfo{title}{Observation of topological flat bands in the kagome semiconductor {Nb}$_3${Cl}$_8$}.
\newblock \emph{\bibinfo{journal}{Nano Letters}} \textbf{\bibinfo{volume}{22}}, \bibinfo{pages}{4596--4602} (\bibinfo{year}{2022}).

\bibitem{Grytsiuk_Mott}
\bibinfo{author}{Grytsiuk, S.}, \bibinfo{author}{Katsnelson, M.~I.}, \bibinfo{author}{van Loon, E. G. C.~P.} \& \bibinfo{author}{R{\"o}sner, M.}
\newblock \bibinfo{title}{{Nb}$_3${Cl}$_8$: A prototypical layered {Mott}--{Hubbard} insulator}.
\newblock \emph{\bibinfo{journal}{npj Quantum Materials}} \textbf{\bibinfo{volume}{9}}, \bibinfo{pages}{8} (\bibinfo{year}{2024}).

\bibitem{Nakamura_HAXPES}
\bibinfo{author}{Nakamura, R.} \emph{et~al.}
\newblock \bibinfo{title}{Charge fluctuations in a cluster {Mott} state: Hard x-ray photoemission study on a breathing kagome magnet {Nb}$_3${Cl}$_8$}.
\newblock \emph{\bibinfo{journal}{Physical Review B}} \textbf{\bibinfo{volume}{110}}, \bibinfo{pages}{L081109} (\bibinfo{year}{2024}).

\bibitem{Gao_flat_Mott}
\bibinfo{author}{Gao, S.}, \bibinfo{author}{Sun, Z.}, \bibinfo{author}{Wang, Y.} \& \bibinfo{author}{et~al.}
\newblock \bibinfo{title}{Discovery of a single-band {Mott} insulator in a van der {Waals} flat band system}.
\newblock \emph{\bibinfo{journal}{Physical Review X}} \textbf{\bibinfo{volume}{13}}, \bibinfo{pages}{041049} (\bibinfo{year}{2023}).

\bibitem{Khan_multiferroic_Nb3Cl8}
\bibinfo{author}{Khan, M.~A.}, \bibinfo{author}{Din, N.~U.}, \bibinfo{author}{Skachkov, D.}, \bibinfo{author}{Englund, D.~R.} \& \bibinfo{author}{Leuenberger, M.~N.}
\newblock \bibinfo{title}{Multiferroic dark excitonic {Mott} insulator in the breathing-kagome lattice material {Nb}$_3${Cl}$_8$}.
\newblock \emph{\bibinfo{journal}{arXiv preprint}}  (\bibinfo{year}{2024}).
\newblock \eprint{2412.13456}.

\bibitem{Skachkov_GW_BSE_NL}
\bibinfo{author}{Skachkov, D.}, \bibinfo{author}{Englund, D.~R.} \& \bibinfo{author}{Leuenberger, M.~N.}
\newblock \bibinfo{title}{Linear and nonlinear optical response based on many-body {GW}--{Bethe}-{Salpeter} and {Kadanoff}--{Baym} approaches for two-dimensional layered semiconductors}.
\newblock \emph{\bibinfo{journal}{npj 2D materials and Applications}}  (\bibinfo{year}{2025}).

\bibitem{Boyd_NLO_review}
\bibinfo{author}{Boyd, R.~W.}
\newblock \emph{\bibinfo{title}{Nonlinear Optics}} (\bibinfo{publisher}{Academic Press}, \bibinfo{address}{San Diego}, \bibinfo{year}{2020}), \bibinfo{edition}{4} edn.

\bibitem{MoS2_SHG_THz}
\bibinfo{author}{Autere, A.} \emph{et~al.}
\newblock \bibinfo{title}{Optical harmonic generation in monolayer group-{VI} transition metal dichalcogenides}.
\newblock \emph{\bibinfo{journal}{Physical Review B}} \textbf{\bibinfo{volume}{98}}, \bibinfo{pages}{115426} (\bibinfo{year}{2018}).

\bibitem{LiNbO3_microresonator}
\bibinfo{author}{Lin, J.} \emph{et~al.}
\newblock \bibinfo{title}{Phase-matched second-harmonic generation in an on-chip {LiNbO}$_3$ microresonator}.
\newblock \emph{\bibinfo{journal}{Physical Review Applied}} \textbf{\bibinfo{volume}{6}}, \bibinfo{pages}{014002} (\bibinfo{year}{2016}).

\bibitem{QE1}
\bibinfo{author}{Giannozzi, P.} \emph{et~al.}
\newblock \bibinfo{title}{Implementation and validation of fully relativistic {GW} calculations: Spin–orbit coupling in molecules, nanocrystals, and solids}.
\newblock \emph{\bibinfo{journal}{J.Phys.: Condens.Matter}} \textbf{\bibinfo{volume}{21}}, \bibinfo{pages}{395502} (\bibinfo{year}{2009}).

\bibitem{QE2}
\bibinfo{author}{Giannozzi, P.} \emph{et~al.}
\newblock \bibinfo{title}{Advanced capabilities for materials modelling with {Quantum ESPRESSO}}.
\newblock \emph{\bibinfo{journal}{J.Phys.: Condens.Matter}} \textbf{\bibinfo{volume}{29}}, \bibinfo{pages}{465901} (\bibinfo{year}{2017}).

\bibitem{Sangalli2019}
\bibinfo{author}{Sangalli, D.} \emph{et~al.}
\newblock \bibinfo{title}{Many-body perturbation theory calculations using the {Yambo} code}.
\newblock \emph{\bibinfo{journal}{J. Phys.: Condens. Matter}} \textbf{\bibinfo{volume}{31}}, \bibinfo{pages}{325902} (\bibinfo{year}{2019}).

\bibitem{Wang2022}
\bibinfo{author}{Wang, Y.} \emph{et~al.}
\newblock \bibinfo{title}{Optical control of high-harmonic generation at the atomic thickness}.
\newblock \emph{\bibinfo{journal}{Nano Lett.}} \textbf{\bibinfo{volume}{22}}, \bibinfo{pages}{8455} (\bibinfo{year}{2022}).

\bibitem{Haraguchi_JPSJ2024}
\bibinfo{author}{Haraguchi, Y.} \& \bibinfo{author}{Yoshimura, K.}
\newblock \bibinfo{title}{Molecular orbital electronic instability in the van der {Waals} kagom{\'e} semiconductor {Nb}$_3${Cl}$_8$: Exploring future directions}.
\newblock \emph{\bibinfo{journal}{Journal of the Physical Society of Japan}} \textbf{\bibinfo{volume}{93}}, \bibinfo{pages}{111002} (\bibinfo{year}{2024}).

\bibitem{Feng_Nb3X8_triferroics}
\bibinfo{author}{Feng, Y.} \& \bibinfo{author}{Yang, Q.}
\newblock \bibinfo{title}{Enabling triferroics coupling in breathing kagome lattice {Nb}$_3${X}$_8$ ({X} = {Cl}, {Br}, {I}) monolayers}.
\newblock \emph{\bibinfo{journal}{Journal of Materials Chemistry C}} \textbf{\bibinfo{volume}{11}}, \bibinfo{pages}{5762--5769} (\bibinfo{year}{2023}).

\bibitem{Regmi_Nb3Br8_flatbands}
\bibinfo{author}{Regmi, S.} \emph{et~al.}
\newblock \bibinfo{title}{Observation of flat and weakly dispersing bands in the van der waals semiconductor {Nb}$_3${Br}$_8$ with breathing kagome lattice}.
\newblock \emph{\bibinfo{journal}{Physical Review B}} \textbf{\bibinfo{volume}{108}}, \bibinfo{pages}{L121404} (\bibinfo{year}{2023}).

\bibitem{Regmi_Nb3I8_flatbands}
\bibinfo{author}{Regmi, S.} \emph{et~al.}
\newblock \bibinfo{title}{Spectroscopic evidence of flat bands in breathing kagome semiconductor {Nb}$_3${I}$_8$}.
\newblock \emph{\bibinfo{journal}{Communications Materials}} \textbf{\bibinfo{volume}{3}}, \bibinfo{pages}{100} (\bibinfo{year}{2022}).

\bibitem{Li_Ta3X8_multiferroics}
\bibinfo{author}{Li, Y.}, \bibinfo{author}{Chen, H.}, \bibinfo{author}{Ren, J.} \emph{et~al.}
\newblock \bibinfo{title}{Two-dimensional multiferroics in a breathing kagome lattice}.
\newblock \emph{\bibinfo{journal}{Physical Review B}} \textbf{\bibinfo{volume}{104}}, \bibinfo{pages}{L060405} (\bibinfo{year}{2021}).

\bibitem{Xing_Ta3I8_triferroic}
\bibinfo{author}{Xing, S.}, \bibinfo{author}{Wang, B.}, \bibinfo{author}{Zhao, T.} \& \bibinfo{author}{Zhou, J.}
\newblock \bibinfo{title}{Independent electrical control of spin and valley degrees in {2D} breathing kagome {Ta}$_3${I}$_8$ with intrinsic triferroicity}.
\newblock \emph{\bibinfo{journal}{The Journal of Physical Chemistry Letters}} \textbf{\bibinfo{volume}{15}}, \bibinfo{pages}{6489--6495} (\bibinfo{year}{2024}).

\bibitem{Lu_Ta3I8_CBV}
\bibinfo{author}{Lu, J.}, \bibinfo{author}{Chen, H.}, \bibinfo{author}{Zhang, Z.}, \bibinfo{author}{Hao, X.} \& \bibinfo{author}{Ren, J.}
\newblock \bibinfo{title}{Chiral breathing-valley locking in two-dimensional kagome lattice {Ta}$_3${I}$_8$}.
\newblock \emph{\bibinfo{journal}{Applied Physics Letters}} \textbf{\bibinfo{volume}{124}}, \bibinfo{pages}{072101} (\bibinfo{year}{2024}).

\bibitem{Sheng_Ta3X8_triplet_EI}
\bibinfo{author}{Sheng, H.} \emph{et~al.}
\newblock \bibinfo{title}{Spin-polarized triplet excitonic insulators in {Ta}$_3${X}$_8$ ({X} = {I}, {Br}) monolayers}.
\newblock \emph{\bibinfo{journal}{arXiv preprint}}  (\bibinfo{year}{2025}).
\newblock \eprint{2506.18686}.

\bibitem{GitHub}
\bibinfo{title}{The software developed and used in this study is openly available on {GitHub}}.
\newblock \bibinfo{howpublished}{\url{https://github.com/LeuenbergerNanoLab}}.

\end{thebibliography}

\end{document}